\documentclass[aps,prd,10pt,notitlepage,nofootinbib,superscriptaddress]{revtex4-1}
\usepackage{amsmath,amssymb,amsfonts}
\usepackage[utf8]{inputenc}
\usepackage{newtxtext,newtxmath}

\usepackage{bbold}
\usepackage{bm}
\usepackage{graphicx}
\usepackage[usenames,dvipsnames]{xcolor}
\usepackage{color}
\usepackage[colorlinks=true,linkcolor=blue,urlcolor=blue,citecolor=blue]{hyperref}

\usepackage{slashed}
\usepackage[english]{babel}
\usepackage{dcolumn}
\usepackage{pifont}
\usepackage{dsfont,mathrsfs}
\usepackage{cancel}
\usepackage{bigints}
\usepackage{accents}
\usepackage{soul}
\usepackage{multirow}

\newcommand{\nn}{\nonumber}

\newcommand{\ensembleaverage}[1]{\left\langle#1\right\rangle}
\newcommand{\MB}[1]{\left|#1\right|}
\newcommand{\FB}[1]{\left(#1\right)}
\newcommand{\SB}[1]{\left\{#1\right\}}
\newcommand{\TB}[1]{\left[#1\right]}

\newcommand{\scrL}{\mathscr{L}}
\newcommand{\scrM}{\mathscr{M}}

\newcommand{\domg}[1]{ \dfrac{d^3{#1}}{(2\pi)^3 2\omega_{#1} }}

\begin{document}
\title{Medium effects on the electrical and Hall conductivities of a hot and magnetized pion gas}

\author{Pallavi Kalikotay}
\email{orionpallavi@gmail.com}
\affiliation{Department of Physics, Kazi Nazrul University, Asansol - 713340, West Bengal, India}

\author{Snigdha Ghosh}
\email{snigdha.physics@gmail.com, snigdha.ghosh@saha.ac.in}
\thanks{(Corresponding Author)}
\affiliation{Saha Institute of Nuclear Physics, 1/AF Bidhannagar, Kolkata - 700064, India}
\affiliation{Government General Degree College at Kharagpur-II, Madpur, Paschim Medinipur - 721149, West Bengal, India}

\author{Nilanjan Chaudhuri}
\email{sovon.nilanjan@gmail.com}
\affiliation{Variable Energy Cyclotron Centre, 1/AF Bidhannagar, Kolkata 700064, India}
\affiliation{Homi Bhabha National Institute, Training School Complex, Anushaktinagar, Mumbai - 400085, India}

\author{Pradip Roy}
\email{pradipk.roy@saha.ac.in}
\affiliation{Saha Institute of Nuclear Physics, 1/AF Bidhannagar, Kolkata - 700064, India}
\affiliation{Homi Bhabha National Institute, Training School Complex, Anushaktinagar, Mumbai - 400085, India}

\author{Sourav Sarkar}
\email{sourav@vecc.gov.in}
\affiliation{Variable Energy Cyclotron Centre, 1/AF Bidhannagar, Kolkata 700064, India}
\affiliation{Homi Bhabha National Institute, Training School Complex, Anushaktinagar, Mumbai - 400085, India}

\begin{abstract}
The electrical and Hall conductivities in a uniform magnetic field are evaluated for an interacting pion gas using the kinetic theory approach within the ambit of relaxation time approximation (RTA). The in-medium cross sections vis-a-vis the relaxation time for $\pi\pi$ scattering are obtained using a one-loop modified thermal propagator for the exchanged $\rho$ and $\sigma$ mesons using thermal field theoretic techniques. For higher values of the magnetic field, a monotonic increase of the electrical conductivity with the temperature is observed. However, for a given temperature the conductivity is found to decrease steadily with magnetic field. The Hall conductivity, at lower values of the magnetic field, is found to decrease with temperature more rapidly than the electrical conductivity, whereas at higher values of the magnetic field, a linear increase is seen. Use of the in-medium scattering cross-section is found to produce a significant effect on the temperature dependence of both electrical and Hall conductivities compared to the case where  vacuum cross-section is used. 
\end{abstract}

\maketitle

\section{Introduction} \label{sec.intro}
The study of strongly interacting matter in the presence of a background magnetic field has significant 
applications in many physical systems (see Ref.~\cite{Kharzeev:2013jha} for a review). In non-central 
Heavy Ion Collisions (HICs), at RHIC and LHC, strong magnetic fields of the order of 
$\sim10^{18}$ Gauss~\cite{Kharzeev:2007jp,Skokov:2009qp} or larger may be  generated due to the collision geometry. 
Note that in natural units, $10^{18} {\rm ~Gauss} \approx m_\pi^2 \approx 0.02~ {\rm GeV}^2$. 
Thus, the fields produced in HICs are comparable to the QCD scale i.e. $eB\approx m_\pi^2 $ and hence it 
can cause noticeable influence on the deconfined medium of quarks and gluons known as quark gluon plasma (QGP). 
This has motivated a large number of investigations on the properties of hot and dense QCD matter in presence 
of background magnetic field in recent times involving several novel and interesting phenomena, such as, 
Chiral Magnetic Effect (CME)~\cite{Kharzeev:2007jp,Fukushima:2008xe,Kharzeev:2009pj,Bali:2011qj}, 
Magnetic Catalysis (MC)~\cite{Shovkovy:2012zn,Gusynin:1994re,Gusynin:1994va,Gusynin:1995nb,Gusynin:1999pq} 
and Inverse Magnetic Catalysis (IMC)~\cite{Preis:2010cq,Preis:2012fh} of dynamical chiral symmetry breaking 
which may cause significant change in the nature of electro-weak~\cite{Elmfors:1998wz,Skalozub:1999jd,Sadooghi:2008yf,Navarro:2010eu}, 
chiral and superconducting phase transitions~\cite{Fayazbakhsh:2010gc,Fayazbakhsh:2010bh,Skokov:2011ib,Fukushima:2012xw}, 
electromagnetically induced superconductivity and superfluidity~\cite{Chernodub:2011mc,Chernodub:2011gs} and many more. 
Besides heavy ion collisions, such magnetic fields of the order of $\sim10^{15}$ Gauss can also be realized on 
the surface of certain compact stars called \textit{magnetars}, while in the interior it is estimated to reach 
magnitudes of the order of  $\sim 10^{18} $ Gauss~\cite{Duncan,Duncan2,Lai}. Cosmological model calculations, 
in fact, predict that during electroweak phase transition in the early universe extremely strong magnetic 
field as high as $ \sim 10^{23}$ Gauss might have been produced~\cite{Vachaspati:1991nm,Campanelli:2013mea}.

The estimation of transport coefficients of relativistic systems in the presence of a magnetic field is 
important in the context of magnetized neutron stars, cosmology and relativistic HICs. 
In case of HICs, transport coefficients such as the shear and the bulk viscosities and the diffusion coefficients 
are essential to describe the hydrodynamical evolution of the matter transiently produced in such collisions. 
In the presence of a magnetic field this evolution is described by magnetohydrodynamics (MHD) which takes 
into account the coupling of the magnetic field to the relativistically evolving fluid in a self-consistent way. 
A good deal of progress has been made in recent times in the evaluation of magnetic field dependent transport 
coefficients such as electrical 
conductivity~\cite{Buividovich:2010tn,Pu:2014fva,Satow:2014lia,Gorbar:2016qfh,Harutyunyan:2016rxm,Kerbikov:2014ofa,Alford:2014doa,Nam:2012sg}, shear viscosity~\cite{Alford:2014doa,Huang:2011dc,Tuchin:2013ie,Finazzo:2016mhm}
heavy quark diffusion constant~\cite{Finazzo:2016mhm,Fukushima:2015wck}, and the jet quenching parameter~\cite{Li:2016bbh}.

One of the most important transport coefficients required in the formulation of MHD is the electrical conductivity. 
Also, from a phenomenological point of view, electrical conductivity is important in the sense that if it is large, 
the created magnetic field in non-central HICs persists for a longer time~\cite{Gursoy:2014aka}. 
Electrical conductivity of a QGP in strong magnetic field has been evaluated in Ref.~\cite{Hattori:2016cnt} where it was
shown that it diverges in the massless limit and is very sensitive to the value of the current quark mass. 
In Ref.~\cite{Feng:2017tsh}, the electrical as well as the Hall conductivities of the QGP have been estimated 
in a strong magnetic field using a kinetic theory approach as well as the Kubo formalism. It is found that, the electrical 
conductivity decreases in the presence of a magnetic field, specially at low temperature. Also in Ref.~\cite{Fukushima:2017lvb}, 
the electrical conductivity of a hot and dense quark matter has been computed in the presence of a magnetic
field using kinetic theory beyond the lowest Landau level (LLL) approximation. It is observed that the transverse electrical 
conductivity is dominated by the Hall conductivity and the parallel conductivity has a nominal dependence on both $T$ and $\mu$.

As it was previously mentioned, the produced magnetic field persists for a longer time if the value of 
electrical conductivity of the medium is large which is a possibility for the case of QGP. 
However, as of now it is a general belief that the value of the magnetic field is quite small in the hadronic
matter (HM) due to the smaller value of the conductivity. As a result, relevant physical quantities calculated 
in HM will have minor modifications as compared to that in quark matter. In order to substantiate this 
conjecture, it is necessary to calculate the $B$-dependent electrical conductivity of HM as accurately as
possible taking into account finite temperature and/or density and magnetic field effects. 
An attempt has been made in Refs.~\cite{Das:2019pqd,Das:2019wjg} 
to calculate the electrical conductivity of Hadron Resonance Gas (HRG) which has been studied using the Relaxation 
Time Approximation (RTA) with a constant cross section whereas in Ref.~\cite{Dash:2020vxk}, electrical conductivity 
along with other transport coefficients of HRG have been computed treating the relaxation time as a free parameter. 
However, the hadronic phase in HICs attains sufficiently 
high temperature ($100$ MeV $\precsim T \precsim 155$ MeV) and/or high (baryon) density (note that the QGP--hadron phase transition occurs nearly at temperature $T_c \simeq 155$ MeV, which is the (pseudo) critical temperature for the chiral phase
transition as obtained in the Lattice QCD calculations~\cite{Bazavov:2014pvz}). 	
Hence in order to obtain a more realistic picture, one should incorporate the 
thermal effects in the cross-sections required to evaluate the transport coefficients which have been ignored in Ref.~\cite{Das:2019wjg}.

In the present work we intend to evaluate the electrical as well as Hall conductivites of a relativistic pion gas using kinetic theory
approach within the RTA where we incorporate the finite temperature effects in the cross-section vis a vis the relaxation time. 
We have chosen to calculate the electrical conductivity of pions as they are the most 
abundant species among the other hadrons produced in the HICs at RHIC and LHC~\cite{Sarkar:2010zza}. 
This type of study is also important as the magnetic field produced in heavy ion collisions is of hadronic scale and hence the
evaluation of transport coefficients of the QGP and the hadronic medium will provide a better insight into the time evolution of strongly
interacting matter in the presence of a background magnetic field.

The article is organized in the following manner. In the next section we evaluate the conductivity tensor using the dissipative 
term obtained from the Boltzmann Transport Equation (BTE) in the presence of external magnetic field employing RTA. 
Sec.~\ref{sec.tau} deals with evaluation of relaxation time of pions in thermal medium using in-medium pionic cross-section. 
In Sec.~\ref{sec.result}, numerical results are shown followed by summary and conclusions in Sec.~\ref{sec.summay}.

\section{ELECTRICAL AND HALL CONDUCTIVITIES FROM KINETIC THEORY} \label{sec.sigma}
Lets us start with the standard expression of the BTE in presence of 
external electromagnetic field~\cite{DeGroot:1980dk} which is satisfied by the 
on-shell single particle phase space distribution function $f_\pm=f_\pm(t,\vec{r},\vec{p})$ of the charged pions ($\pi^\pm$) as 
\begin{eqnarray}
\frac{\partial f_\pm}{\partial t} + \vec{v}\cdot\frac{\partial f_\pm}{\partial \vec{r}} \pm q\TB{\vec{E} + (\vec{v}  \times \vec{H})}\cdot\frac{\partial f_\pm}{\partial \vec{p}} = C[f_\pm] \label{BTE}
\end{eqnarray} 
where, $q$ is the charge of a proton, $\vec{v}=\vec{p}/\omega_p$ is the velocity, $\omega_p=\sqrt{\vec{p}^2+m^2}$ is the energy, $\vec{E}$ is the electric field, 
$\vec{H}$ is the magnetic field and $C[f]$ denotes the collision kernel. We note that, the equilibrium Bose-Einstein distribution 
function $f_0=f_0(\omega_p)$ for which $C[f_0]=0$  is given by
\begin{eqnarray}
f_0 (\omega_{p}) = \frac{1}{e^{\omega_p/T} - 1} \label{eq_dist}
\end{eqnarray}
where $T$ is temperature.

Few comments on the use of the classical dispersion relation $\omega_p = \sqrt{\vec{p}^2+m^2}$ for the pions are in order here. It is well known that in the presence of an external magnetic field the  momentum states of the charged pions will be Landau quantized and their classical dispersion relation $\omega_p = \sqrt{\vec{p}^2+m^2}$ with continuous transverse momentum modifies to 
\begin{eqnarray}
\omega_{pl} = \sqrt{p_z^2 + (2l+1)qH + m^2} \label{eq.LQ}
\end{eqnarray}
where $l$ is the Landau level. However, in this work we have ignored the Landau quantization (LQ) in the calculation of conductivity assuming the magnetic field to be weak. For low values of the external magnetic field, the Landau levels become closely spaced such that the continuum approximation of the transverse momentum of pions holds good. Moreover, we will be using the RTA, where the pion distribution function is assumed to be slightly away from equilibrium which allows linearization of the BTE. The use of RTA implies that the external magnetic field cannot be too high. Finally, the magnitude of external magnetic field in hadronic phase of HIC is usually small which in turn justifies the use of weak field approximation in our calculation. A more quantitative analysis on the validity of the continuum approximation will be performed later in Sec.~\ref{sec.result}.

When the system is out of equilibrium, the dissipative processes within the system try to bring it back to equilibrium. 
Let us consider the system to be slightly away from equilibrium which is characterized by the non-equilibrium distribution function $f_\pm=f_0+\delta f_\pm$ 
with $\delta f_\pm = -\phi_\pm \dfrac{\partial f_0}{\partial\omega_p} \ll f_0$. As $\delta f_\pm$ is small, the BTE can be linearized.

In order to solve Eq.~\eqref{BTE} we treat the right hand side of Eq.~\eqref{BTE} using the RTA and consider 
only the $2\to2$ scattering process $k + p\to k' + p'$. In RTA, the test particle with momentum $p$ is considered to be out of equilibrium whereas 
the remaining three particles with momenta $k$, $k'$ and $p'$ are in equilibrium. Thus, in the RTA the collision integral in Eq.~\eqref{BTE} reduces to~\cite{DeGroot:1980dk} 
\begin{eqnarray}
C[f_\pm]= \frac{\delta f_\pm}{\tau} = -\frac{\phi_\pm}{\tau} \frac{\partial f_0}{\partial \omega_p} \label{C_RTA}
\end{eqnarray}
where $\tau$ is the relaxation time. Substituting Eq.~\eqref{C_RTA} into Eq.~\eqref{BTE} yields
\begin{eqnarray}
\frac{\partial f_\pm}{\partial t} + \vec{v}\cdot\frac{\partial f_\pm}{\partial \vec{r}} 
\pm q\TB{\vec{E} + (\vec{v}  \times \vec{H})} \cdot \frac{\partial f_\pm}{\partial \vec{p}} 
= -\frac{\phi_\pm}{\tau} \frac{\partial f_0}{\partial \omega_p} \label{BTE_phi}
\end{eqnarray} 
Since we are dealing with an uniform and static medium, both $f_\pm$ and $f_0$ are independent of time and space. 
Also the electric field under consideration is very small. Thus Eq.~(\ref{BTE_phi}) reduces to
\begin{eqnarray}
\pm q\vec{E}\cdot\vec{v}\frac{\partial f_0}{\partial \omega_p} \pm q(\vec{v}\times\vec{H}) \cdot \frac{\partial \phi_\pm}{\partial \vec{p}} \FB{\frac{\partial f_0}{\partial \omega_p}}
= -\frac{\phi_\pm}{\tau} \frac{\partial f_0}{\partial \omega_p} \label{BTE_final}
\end{eqnarray}
In order to solve for the electrical and Hall conductivities, we take the following ansatz for the functional form of $\phi_\pm$~\cite{Harutyunyan:2016rxm} as
\begin{eqnarray}
\phi_\pm= \vec{p}\cdot \vec{\Xi}_\pm(\omega_p) 
\label{phi}
\end{eqnarray}
where the vector $\vec{\Xi}_\pm$ contains information of the dissipation produced due to electric and magnetic fields and 
can be expressed most generally as 
\begin{eqnarray}
\vec{\Xi}_\pm  =  \alpha_\pm \hat{e} + \beta_\pm \hat{h} + \gamma_\pm \FB{\hat{e}\times \hat{h}} 
\label{Xi_equation}
\end{eqnarray}
where, $\hat{e}=\vec{E}/|\vec{E}|$ and $\hat{h}=\vec{H}/|\vec{H}|$ are the unit vectors along the directions of electric and magnetic fields respectively. 
Substituting Eq.~\eqref{phi} and Eq.~\eqref{Xi_equation} into Eq.~\eqref{BTE_final}, we get after some simplification
\begin{eqnarray}
\pm \frac{q|\vec{E}|}{\omega_p}(\hat{e}\cdot \vec{p}) \pm \alpha_\pm \frac{q|\vec{H}|}{\omega_p} (\hat{e}\times \hat{h})\cdot \vec{p} \pm \gamma_\pm \frac{q|\vec{H}|}{\omega_p} (\hat{e}\cdot \vec{p}) 
\pm \gamma_\pm \frac{q|\vec{H}|}{\omega_p}(\hat{h}\cdot \vec{p}) (\hat{h}\cdot \hat{e}) \nn \\
= -\alpha_\pm (\hat{e}\cdot \vec{p})\tau^{-1} 
- \beta_\pm (\hat{h}\cdot \vec{p}) \tau^{-1} - \gamma_\pm (\hat{e}\times\hat{h}) \cdot \vec{p}\tau^{-1}.
 \label{eq.abc}
\end{eqnarray}
Comparing the coefficients of $\hat{e}\cdot\vec{p}$, $\hat{e}\times\hat{h}$ and $\hat{h}\cdot\vec{p}$ on both sides of Eq.~\eqref{eq.abc}, we obtain
\begin{eqnarray}
\alpha_\pm &=& \pm \frac{q|\vec{E}|}{\omega_p} \frac{\tau}{1+\omega_c^2\tau^2}, \label{eq.a}\\
\frac{\beta_\pm}{\alpha_\pm} &=& -(\omega_c\tau)^2 (\hat{h}\cdot \hat{e}), \label{eq.b} \\
\frac{\gamma_\pm}{\alpha_\pm} &=& -\omega_c \tau \label{eq.c}
\end{eqnarray}
where $\omega_c = |q\vec{H}|/\omega_p$ is the cyclotron frequency. 
Using Eqs.~\eqref{eq.a}-\eqref{eq.c} in Eq.~\eqref{Xi_equation}, we can now obtain the vector $\vec{\Xi}_\pm$ which in turn is used to get $\phi_\pm$ from Eq.~\eqref{phi} as
\begin{eqnarray}
\phi_\pm = \alpha_\pm \omega_p \vec{v} \cdot \TB{1 + (\omega_c \tau)^2 (\hat{e} \cdot \hat{h})\hat{h} - (\omega_c\tau) (\hat{e}\times \hat{h}) }  
= \pm \frac{q\tau}{1+(\omega_c \tau)^2}  v^i \TB{\delta^{ij} - \omega_c \tau \epsilon^{ijk} h^k + (\omega_c\tau)^2 h^i h^j } E^j
\label{phi_final}
\end{eqnarray}
where the Einstein summation convention has been used.

In order to extract the electrical and Hall conductivities from $\phi_\pm$, we first note that the macroscopic electrical current density $j^i$ is given by
\begin{eqnarray}
j^i = \sigma^{ij} E^j= \!\!\int\!\! \frac{d^3p}{(2\pi)^3} v^i  q(\phi_+ -\phi_-) \FB{ \frac{\partial f_0}{\partial \omega_p} }.
\label{j}
\end{eqnarray}
In Eq.~\eqref{j}, $\sigma^{ij}$ is the conductivity tensor. Now substitution of Eq.~\eqref{phi_final} into Eq.~\eqref{j} yields
\begin{eqnarray}
\sigma^{ij}=~\delta^{ij}\sigma_0 -\epsilon^{ijk} h^k \sigma_1  + h^i h^j \sigma_2 
\end{eqnarray}
where
\begin{eqnarray}
\sigma_0 &=& \frac{g q^2}{3T} \!\!\int \!\!\frac{d^3 p}{(2\pi)^3} \frac{\vec{p}^2}{\omega_p^2} \frac{\tau}{1+ (\omega_c \tau)^2} f_0(\omega_p)\SB{1+f_0(\omega_p)}, \label{sigma_0} \\
\sigma_1 &=& \frac{g q^2}{3T}\!\! \int\!\! \frac{d^3 p}{(2\pi)^3} \frac{\vec{p}^2}{\omega_p^2} \frac{\tau (\omega_c \tau)}{1+ (\omega_c \tau)^2} f_0(\omega_p)\SB{1+f_0(\omega_p)}, \label{sigma_1} \\
\sigma_2 &=& \frac{g q^2}{3T} \!\!\int\!\! \frac{d^3 p}{(2\pi)^3} \frac{\vec{p}^2}{\omega_p^2} \frac{\tau (\omega_c \tau)^2}{1+ (\omega_c \tau)^2} f_0(\omega_p)\SB{1+f_0(\omega_p)} \label{sigma_2}
\end{eqnarray}
in which $g=2$ is the degeneracy of charged pions in the gas since only the charged pions $\pi^+$ and $\pi^-$ participate in the charge conduction. 
$\sigma_0$  is the electrical conductivity in presence of the magnetic field, $\sigma_1$ is the Hall conductivity and 
$\sigma_0+\sigma_2$ is the electrical conductivity in absence of external magnetic field. 
In compact notation, Eqs.~\eqref{sigma_0}-\eqref{sigma_2} can be written as
\begin{eqnarray}
\sigma_n = \frac{g q^2}{3T} \!\! \int \!\! \frac{d^3 p}{(2\pi)^3} \frac{\vec{p}^2}{\omega_p^2} \frac{\tau (\omega_c \tau)^n}{1+ (\omega_c \tau)^2} f_0(\omega_p)\SB{1+f_0(\omega_p)} \label{sigma_n}
~~~;~~~ n=0,1,2.
\end{eqnarray}

\section{THE RELAXATION TIME IN THE MEDIUM} \label{sec.tau}
The relaxation time $\tau$ which appears in Eq.~\eqref{sigma_n} is the key dynamical input and for the $2\to2$ 
process ($\pi (k) + \pi (p) \to \pi(k') + \pi(p'))$ it is given by~\cite{Prakash:1993bt}
\begin{eqnarray}
\TB{\tau(p)}^{-1} &=& \frac{g'}{4 \omega_{p}}\! \int\!\!\!\!\int\!\!\!\!\int\!\! \domg{k} \domg{k'} \domg{p'}(2\pi)^4 \delta ^4 \FB{k + p - k' -p' }
\MB{ \scrM }^2   \frac{f_0(\omega_k) \left\{1+f_0(\omega_{k'})\right\}  \left\{1+f_0(\omega_{p'}) \right\} } { \left\{ 1+ f_0(\omega_p)\right\}} \nn \\ 
\end{eqnarray}
where the total pion degeneracy $g'=3$ on account of the fact that both the charged and neutral pions ($\pi^\pm$ and $\pi^0$) participate in the 
scattering processes via effective strong interaction. Considering $ f_0(\omega_{p'}) \simeq f_0(\omega_{p}) $ 
and $ f_0(\omega_{k'}) \simeq f_0(\omega_{k}) $~\cite{Prakash:1993bt}, 
we can integrate over the momenta of the final state particles $k'$ and $p'$ respectively obtaining
\begin{equation}
\TB{\tau(p)}^{-1} = \frac{g'}{2} \!\int\!\!\! \frac{d^3k}{(2\pi)^3} \FB{\sigma v_\text{rel}} f_0(\omega_k)\SB{1 + f_0(\omega_k)}
\label{tau_k}
\end{equation} 
where, $\sigma$ is the total cross section for the $2\to2$ scattering process and $v_\text{rel}=\frac{1}{2\omega_k\omega_p}\lambda^{\frac{1}{2}}\FB{(\omega_k+\omega_p)^2,m^2,m^2}$ is the relative velocity of the 
initial state particles in which $\lambda(x,y,z)=x^2+y^2+z^2-2xy-2yz-2zx$ is the K\"all\'en function.

Let us now proceed to calculate $\pi\pi$ cross section $\sigma$ in a thermal medium. The effective interaction of pions with 
the vector meson $\rho$ and scalar meson $\sigma$ is given by the following Lagrangian (density)~\cite{Krehl:1999km}
\begin{eqnarray}
\scrL_\text{int} = g_{\rho\pi\pi}\vec{\rho}_\mu\cdot\vec{\pi}\times\partial^\mu\vec{\pi} + 
\frac{1}{2}g_{\sigma\pi\pi}m_\sigma\vec{\pi}\cdot\vec{\pi}\sigma 
\label{eq.lagrangian}  
\end{eqnarray}
where the coupling constants $g_{\rho\pi\pi}=6.05$ and $g_{\sigma\pi\pi}=2.5$ have been obtained from the experimental decay widths of $\rho$ and $\sigma$ mesons~\cite{Krehl:1999km,Mallik:2016anp}. 
It is now convenient to use the isospin basis so that the invariant amplitudes $\scrM_I$ for the particular isospin channel with total isospin $I$ are given by~\cite{Mitra:2012jq} 
\begin{eqnarray}
\scrM_2 &=& g_{\rho\pi\pi}^2\left[-\left(\frac{s-u}{t-m_\rho^2}\right)-
\left(\frac{s-t}{u-m_\rho^2}\right) \right] +4g_{\sigma\pi\pi}^2\left[\frac{1}{t-m_\sigma^2} + \frac{1}{u- m_\sigma^2}\right], \\
\scrM_1 &=& g_{\rho\pi\pi}^2\left[ 2\left(\frac{t-u}{s-m_\rho^2 - \Pi_\rho}\right)+\left(\frac{s-u}{t-m_\rho^2}\right)
-\left(\frac{s-t}{u-m_\rho^2}\right)\right] + 4g_{\sigma\pi\pi}^2\left[\frac{1}{t-m_\sigma ^2}- \frac{1}{u - m_\sigma^2}\right], \\
\scrM_0 &=& g_{\rho\pi\pi}^2\left[2\left(\frac{s-u}{t- m_\rho^2}\right)+2\left(\frac{s-t}{u-m_\rho^2}\right)\right]
+4g_{\sigma\pi\pi}^2\left[\frac{3}{s-m_\sigma^2 - \Pi_\sigma} + \frac{1}{t- m_\sigma^2} + \frac{1}{u- m_\sigma^2}\right]
\end{eqnarray}
where $s=(k+p)^2$, $t=(k-k')^2$ and $u=(k-p')^2$ are the Mandelstam variables and the vacuum propagators for the $\rho$ and $\sigma$ in their respective s-channels 
have been replaced by the complete interacting (dressed) propagators obtained from a Dyson-Schwinger sum involving the one-loop in-medium self energies of $\rho$ and $\sigma$ 
denoted by $\Pi_\rho$ and $\Pi_\sigma$ respectively.

The one-loop self energies $\Pi_h(q)$ of $h\in\{\rho,\sigma\}$ at finite temperature can be calculated using the standard techniques 
of Real Time Formalism (RTF) of finite temperature field theory~\cite{Mallik:2016anp,Bellac:2011kqa}. 
Contributions to $\Pi_h$ come from different loop graphs containing other mesons ($i,j$). 
The $\rho$ self energy consists of $\{i,j\}$ = $\{\pi,\pi\}$, $\{\pi,\omega\}$, $\{\pi,h_1\}$ and $\{\pi,a_1\}$ 
loops whereas the $\sigma$ self energy has contribution from only $\{i,j\}$ = $\{\pi,\pi\}$ loop. 
In a general notation, the real part of the self energy reads
\begin{eqnarray}
\text{Re}~\Pi_h(q) &=& \sum_{ \{\pi,j\} \in \{\text{loops}\}}  
\int\!\!\!\frac{d^3k}{(2\pi)^3} \frac{1}{2 \omega_{k}\omega_{p_j}}\mathcal{P}\left[
\left(\frac{ f_0(\omega_{k}) \omega_{p_j} \mathcal{N}_{\pi j}^h(k^0=\omega_{k})}{(q_0-\omega_{k})^2-\omega_{p_j}^2}\right) +
\left(\frac{ f_0(\omega_{k}) \omega_{p_j} \mathcal{N}_{\pi j}^h(k^0=-\omega_{k})}{(q_0+\omega_{k})^2-\omega_{p_j}^2}\right)\right. \nn\\ 
&&\hspace{3cm}\left. +\left(\frac{ f_0(\omega_{p_j})  \omega_{k} \mathcal{N}_{\pi j}^h (k^0=q_0-\omega_{p_j})}{(q_0-\omega_{p_j})^2-\omega_{k}^2}\right) 
+ \left(\frac{ f_0(\omega_{p_j}) \omega_{k} \mathcal{N}_{\pi j}^h (k^0=q_0+\omega_{p_j})}{(q_0+\omega_{p_j})^2-\omega_{k}^2}\right) \right] \label{eq.repi}
\end{eqnarray}
whereas the imaginary part is
\begin{eqnarray}
\text{Im}~\Pi_h(q) &=& -\pi\text{sign}(q_0) \!\!\!\! \sum_{ \{\pi,j\} \in \{\text{loops}\}} 
\int\!\!\!\frac{d^3k}{(2\pi)^3}\frac{1}{4 \omega_k\omega_{p_j}} \nn \\
&& \times \Big[ \SB{1 + f_0(\omega_k)+f_0(\omega_{p_j})} \Big\{ \mathcal{N}_{\pi j}^h(k^0=\omega_k) \delta(q_0-\omega_k-\omega_{p_j}) 
 - \mathcal{N}_{\pi j}^h(k^0=-\omega_k) \delta(q_0+\omega_k+\omega_{p_j}) \Big\} \nn \\ &&
-\SB{f_0(\omega_k)-f_0(\omega_{p_j})} \Big\{ \mathcal{N}_{\pi j}^h(k^0=\omega_k)\delta(q_0-\omega_k+\omega_{p_j})  - \mathcal{N}_{\pi j}^h(k^0=-\omega_k)\delta(q_0+\omega_k-\omega_{p_j}) 
\Big\} \Big].
\label{eq.impi}
\end{eqnarray}
where, $\omega_{p_j} = \sqrt{\vec{p}_j^2+m_j^2}$ with $p_j=(q-k)$. Although the self energy of $\rho$ contains additional Lorentz indices, 
in this work we have used the polarization averaged self energy of $\rho$ in the expression of the invariant amplitude. 
It is also to be noted that, corresponding to the loop graphs $\{\pi,h_1\}$ and $\{\pi,a_1\}$ contributing to the self energy of $\rho$, 
$\Pi_\rho$ has been convoluted with the vacuum spectral functions of the unstable mesons $h_1$ and $a_1$ due to their mass 
uncertainties~\cite{Sarkar:2004jh} due to large width. 
The detailed expressions of $\mathcal{N}_{ij}^\rho$ and $\mathcal{N}_{ij}^\sigma$ can be found in Ref.~\cite{Mitra:2012jq}. 
In the expression of the imaginary part of the self energy, the four terms containing the Dirac delta functions  
correspond to different physical processes like decay and scattering leading to the absorption of the meson $h$ in the thermal medium.

The isospin averaged invariant amplitude 
\begin{eqnarray}
\overline{\MB{\scrM}^2} &=& \sum_{I}(2I + 1)\MB{\scrM_I}^2 \Big/ \sum_{I} (2I+1)~,
\end{eqnarray}
is used to obtain the total $\pi\pi\to\pi\pi$ cross section from 
\begin{eqnarray}
\sigma(s)=\frac{1}{64\pi^2 s}\int \!\! d\Omega~\overline{\MB{\scrM}^2}.
\end{eqnarray}
\begin{figure}[h]
	\begin{center}
		\includegraphics[angle=-90,scale=0.34]{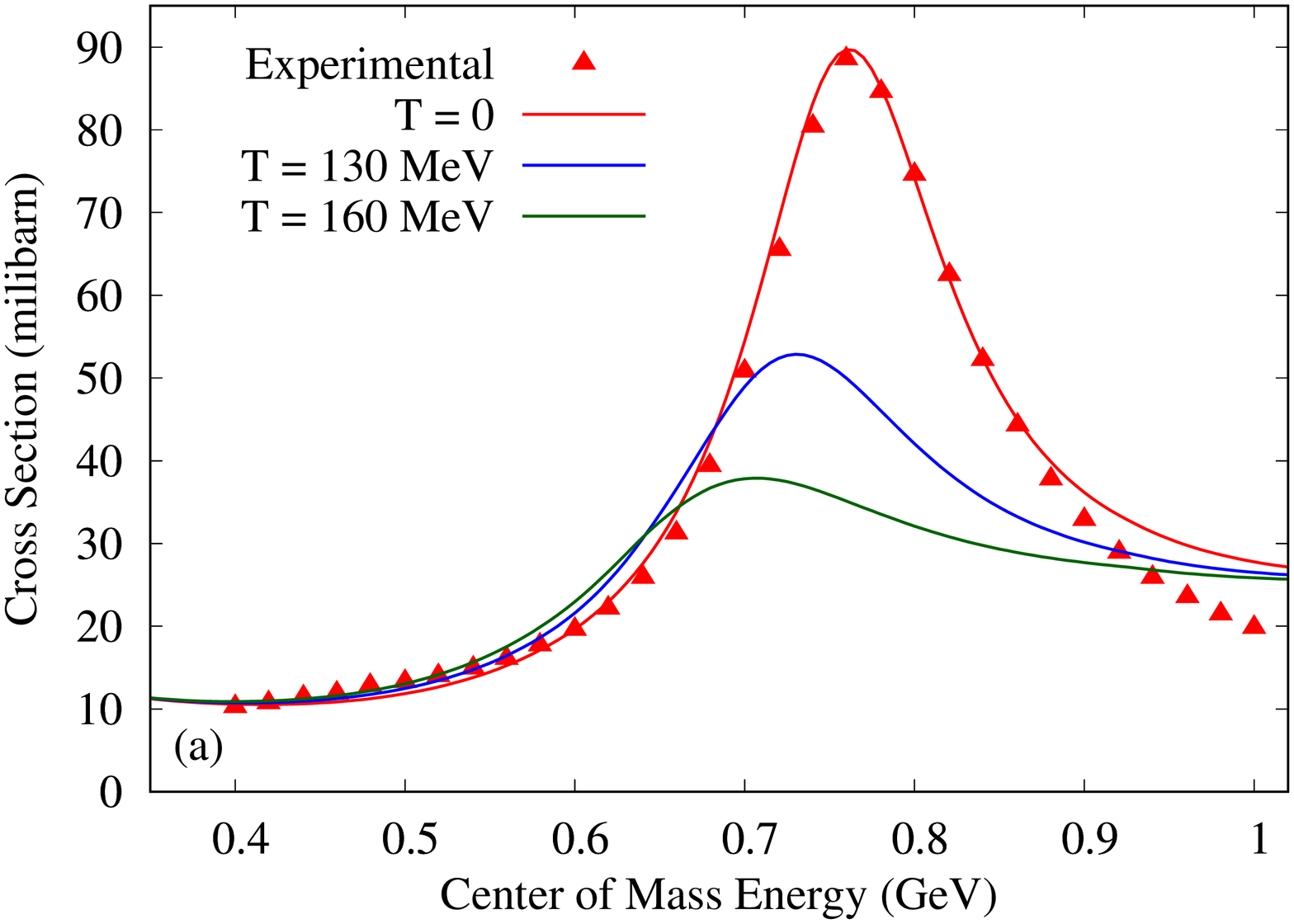}  \includegraphics[angle=-90,scale=0.34]{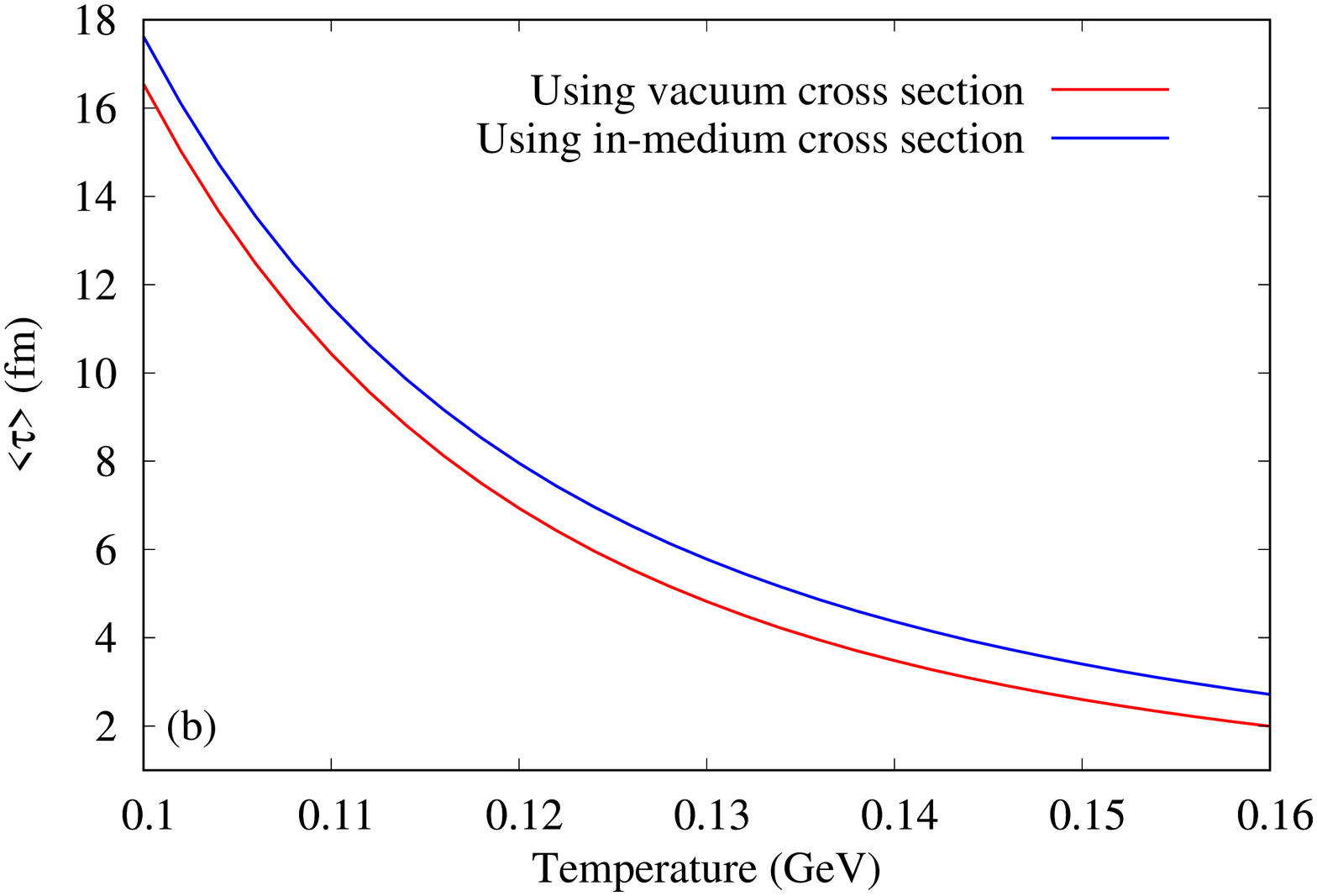}
		\includegraphics[angle=-90,scale=0.34]{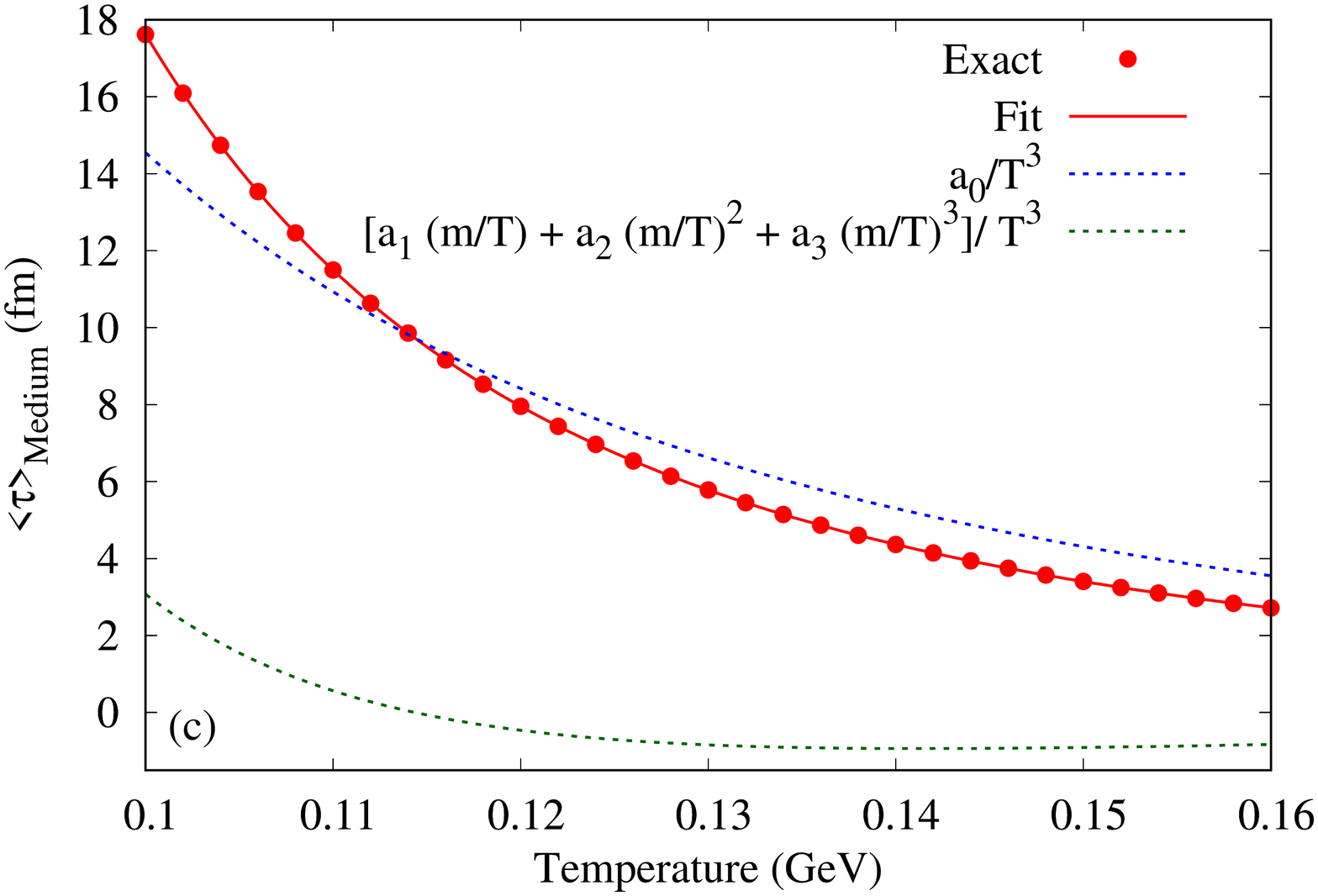}
	\end{center}
	\caption{(Color Online) (a) The variation of isospin averaged total $\pi\pi\to\pi\pi$ cross section as a function of centre of mass energy for different temperatures. 
		The experimental data is taken from Ref.~\cite{Prakash:1993bt}. (b) The variation of average relaxation time $\ensembleaverage{\tau}$ of pions as a function of temperature calculated using the vacuum and in-medium cross sections. (c) The in-medium $\ensembleaverage{\tau}$ as a function of $T$ fitted with a polynomial funtion of the form $\sum\limits_{i=0}^{3} a_i\FB{\frac{m}{T}}^i \frac{1}{T^3}$ with $a_0 = 0.0145$ fm-GeV$^{3}$, $a_1 = -0.0109$ fm-GeV$^{3}$, $a_2 = 0.0058$ fm-GeV$^{3}$ and $a_3 = 0.0026$ fm-GeV$^{3}$.}
	\label{fig.xsection}
\end{figure}
\section{RESULTS \& DISCUSSIONS} \label{sec.result}
We begin this section by showing the in-medium $\pi\pi\to\pi\pi$ total cross section as a function of center of mass energy in Fig.~\ref{fig.xsection}(a) for different temperatures. 
At $T=0$, the cross section obtained using the vacuum self energies of $\rho$ and $\sigma$ mesons is seen to agree with the experimental data~\cite{Prakash:1993bt} shown with red triangles. 
With the increase in temperature, the imaginary part of the self energy increases owing to the in-medium broadening of the resonance spectral functions. Physically, it 
corresponds to the increase in annihilation probabilities (due to decay and scattering) of $\rho$ and $\sigma$ in the thermal bath. 
This in-medium spectral-broadening of $\rho$ and $\sigma$ in turn makes substantial suppression in the cross section at high temperature as can be noticed from the figure.

Next, in Fig.~\ref{fig.xsection}(b) we have shown the variation of the average relaxation time $\ensembleaverage{\tau}$ of pions with temperature evaluated using 
vacuum and in-medium cross sections. Note that, the momentum averaged relaxation time $\ensembleaverage{\tau}$ is obtained from the relation 
\begin{eqnarray}
\ensembleaverage{\tau} = \int\!\! d^3p \tau(p) f(\omega_p) \Big/ \int\!\! d^3p f(\omega_p). \label{eq.tauav}
\end{eqnarray}
It is seen that the relaxation time obtained using the in-medium cross section is 
always greater than the same calculated using the vacuum cross section which is also obvious from Eq.~\eqref{tau_k}. Since, the 
in-medium cross section is suppressed with respect to the vacuum cross section, $\ensembleaverage{\tau}_\text{Vacuum}$ comes out to be less than $\ensembleaverage{\tau}_\text{Medium}$. 
In order to extract the leading behavior of $\ensembleaverage{\tau}$ as a function of temperature, we fit the in-medium relaxation time $\ensembleaverage{\tau}_\text{Medium}$ with a polynomial function of the form $\sum\limits_{i=0}^{3} a_i\FB{\frac{m}{T}}^i \frac{1}{T^3}$. This is shown in Fig.~\ref{fig.xsection}(c) where we have plotted the fitted function along with $ \FB{\dfrac{a_0}{T^3}}$ and $\sum\limits_{i=1}^{3} a_i\FB{\frac{m}{T}}^i \frac{1}{T^3}$ separately to understand the leading behavior. It is easy to check from Eqs.~\eqref{tau_k} and \eqref{eq.tauav} that $a_i$'s are dimensionfull quantities and have the dimensions of the inverse of the cross section, $[\sigma^{-1}]$. It is clearly seen that the leading behavior is well represented by the first term in the fitting function. This can be explained by considering $\ensembleaverage{\tau}\sim 1/(n \sigma)$ where $n$ is the pion density which goes as $n \sim T^3$ in the massless limit and $\sigma$ is the ($T$-independent) cross section. The observed deviation from the $1/T^3$ behavior of $\ensembleaverage{\tau}$ at lower and higher temperatures is quite understandable and may be attributed to several factors. Most important among these is the contribution coming from the phase space integrals due to the non-zero pion mass  which contain higher inverse powers of $T$. The $T$-dependence of the cross-section can also make a contribution. However, for purposes of discussion $\ensembleaverage{\tau}$ may well be taken to go as $1/T^3$ in the relevant temperature range  and the deviations therefrom will not affect the conclusions significantly.
\begin{figure}[h]
	\begin{center}
		\includegraphics[angle=-90,scale=0.31]{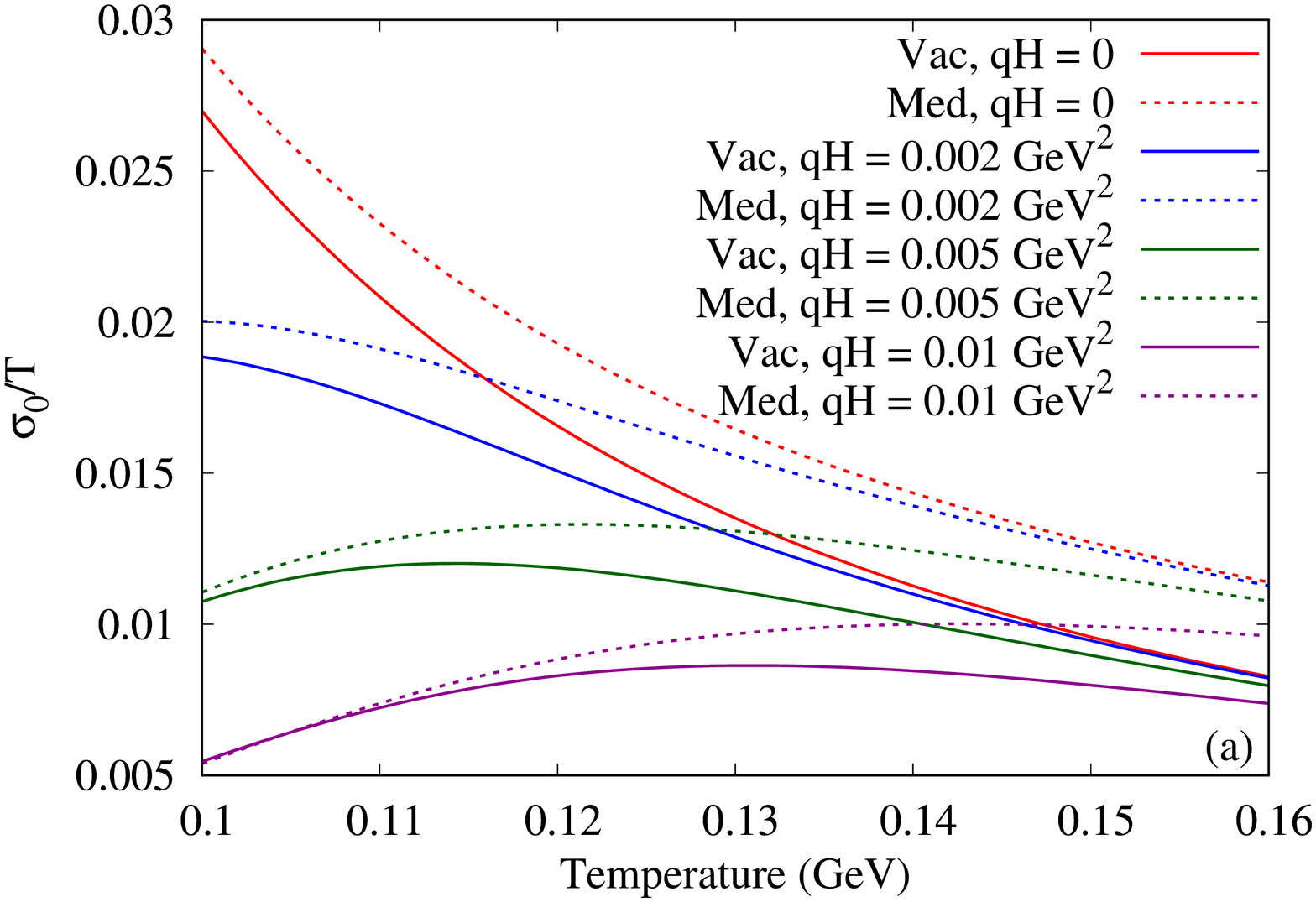}  \includegraphics[angle=-90,scale=0.31]{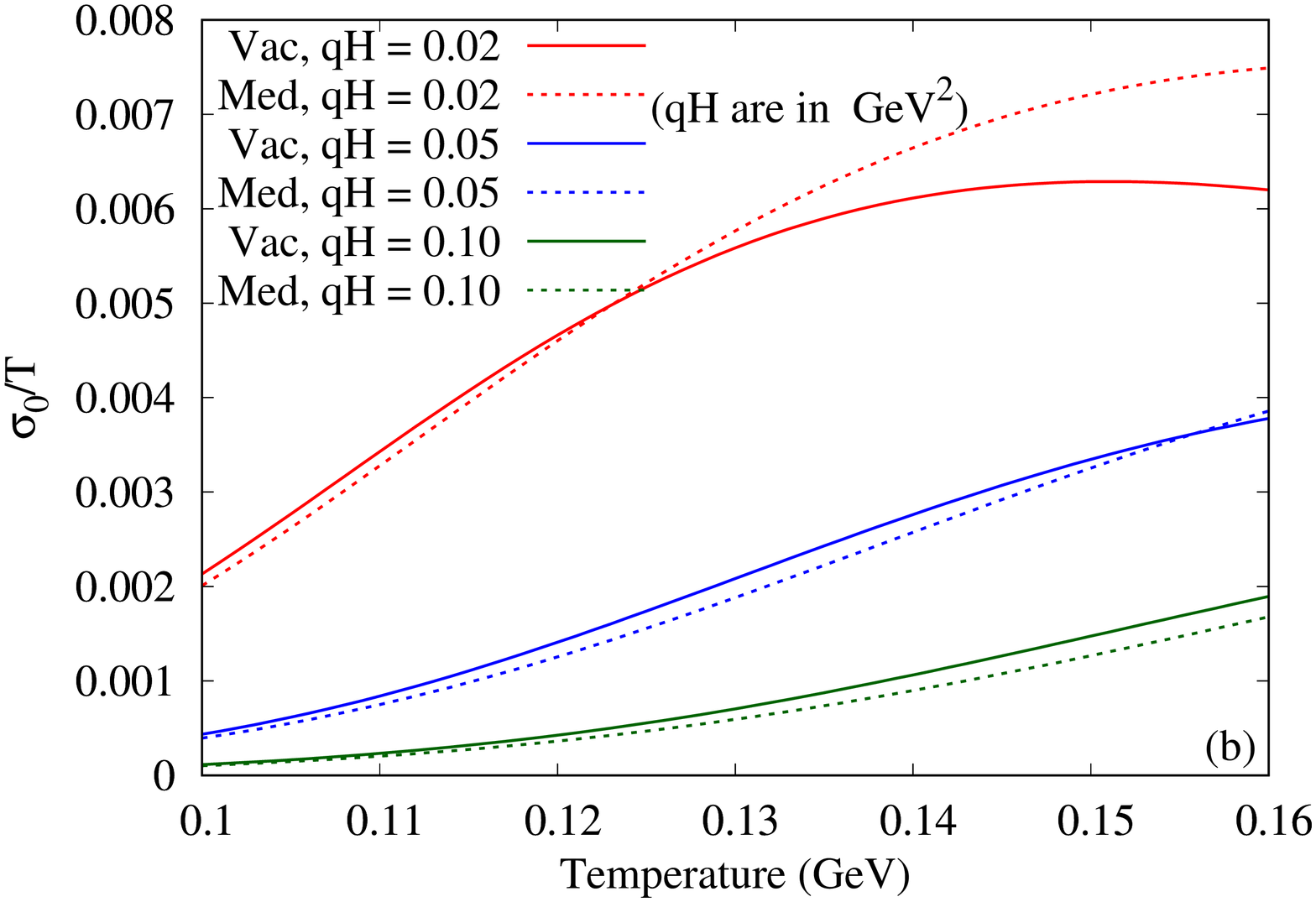}
		\includegraphics[angle=-90,scale=0.31]{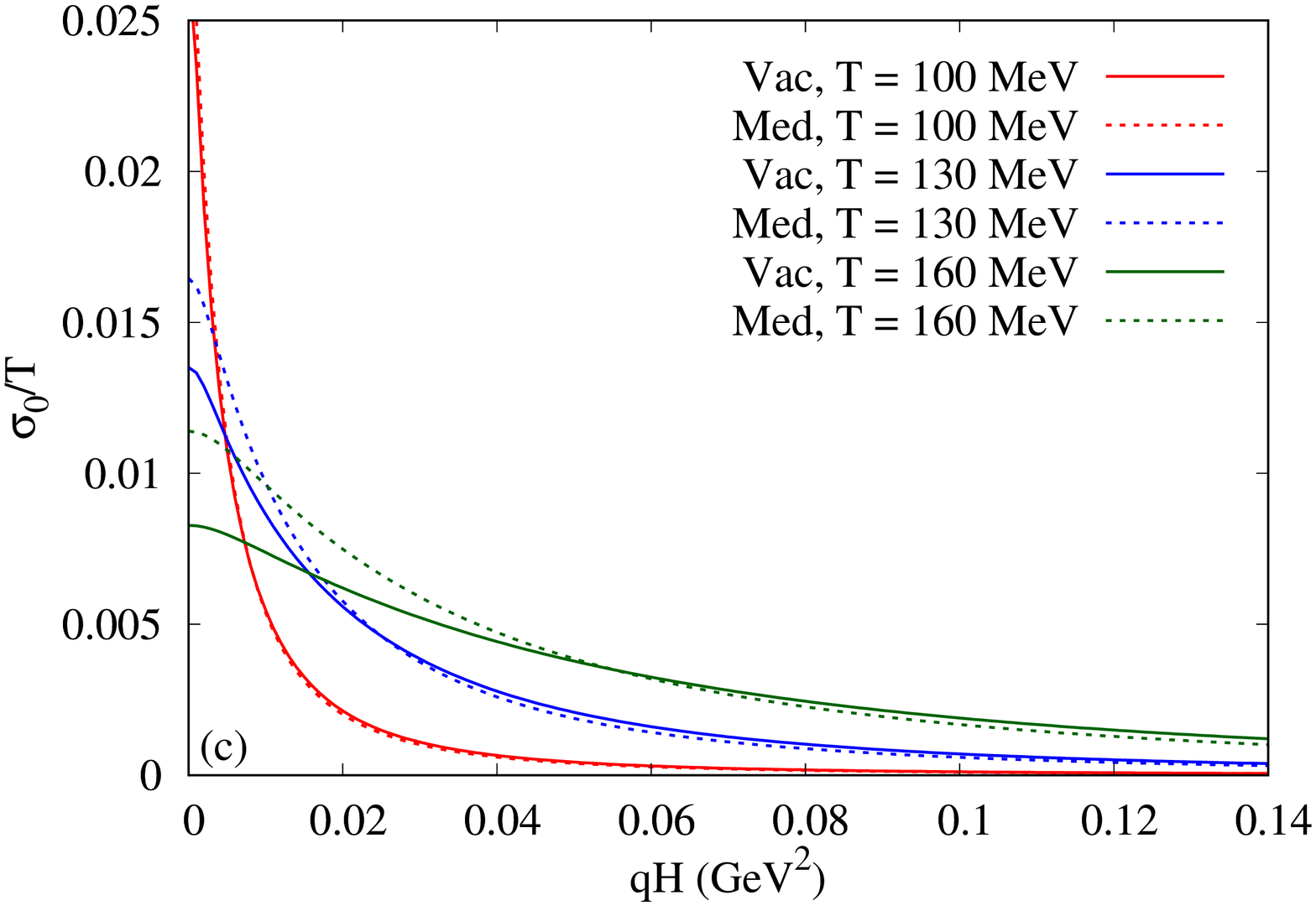}
	\end{center}
	\caption{(Color Online) The variation of $\sigma_0/T$ (a)-(b) as a function of temperature for different values of magnetic field strength and (c) as a function of magnetic field for different values of temperature. The solid and dashed curves correspond to the estimations of $\sigma_0/T$ using vacuum and in-medium cross sections respectively.}
	\label{fig.sig0}
\end{figure}

The variation of $\sigma_0/T$ as a function of temperature is shown in Figs.~\ref{fig.sig0}(a) and \ref{fig.sig0}(b) for different values of magnetic field  using both the vacuum and in-medium cross sections. To understand the behavior of $\sigma_0/T$ with temperature, we first note, from Eq.~\eqref{sigma_0}, that the temperature dependence 
of $\sigma_0/T$ roughly comes from $\sigma_0/T \sim \frac{\tau T}{1+(\omega_c\tau)^2}$. At lower values of the magnetic field, $\omega_c\tau \ll 1$ so that the temperature dependence of $\tau T$ would dictate the temperature dependence of  $\sigma_0/T$. 
We have discussed the $T$-dependence of the average relaxation time earlier by fitting a simple function and argued that $\ensembleaverage{\tau}\propto\,1/T^3$ is a good approximation in the relevant temperature range.
Thus, at lower values of 
magnetic field, we expect $\sigma_0/T\sim 1/T^2$ which is quite compatible with Fig.~\ref{fig.sig0}(a). The situation is reversed at higher values of the external magnetic field 
for which $\omega_c\tau \gg 1$ and consequently the temperature dependence of $\sigma_0/T$ approximately comes from $\sigma_0/T \sim \frac{T}{\tau}\sim T^4$. We thus expect a monotonically increasing 
trend of $\sigma_0/T$ with temperature at higher values of magnetic field as can be noticed in Fig.~\ref{fig.sig0}(b). At intermediate values of the magnetic field we
observe a non-monotonic behavior of $\sigma_0/T$ with temperature. 
In Fig.~\ref{fig.sig0}(c), $\sigma_0/T$ has been plotted as a function of external magnetic field for different temperatures. 
Unlike the temperature dependence, $\sigma_0/T$ has a trivial magnetic field dependence as $\sigma_0/T \sim \frac{1}{1+(\omega_c\tau)^2}$. 
With the increase in magnetic field values, the cyclotron frequency $\omega_c$ increases linearly so that a monotonically decreasing trend 
of $\sigma_0/T$ with external magnetic field is visible in Fig.~\ref{fig.sig0}(c).
The effect of in-medium cross section on $\sigma_0/T$ can be understood similarly from the $\tau$ dependence of $\sigma_0/T$.  As already argued, at a given temperature, for lower values of 
magnetic field, $\sigma_0/T\sim \tau$ whereas for higher values of magnetic field $\sigma_0/T\sim 1/\tau$. Since, the relaxation time is larger for the in-medium cross section, 
it is obvious that the use of in-medium cross section instead of the vacuum cross section will increase (decrease) $\sigma_0/T$ for lower (higher) values of the external magnetic field. This is clearly observed in Figs.~\ref{fig.sig0}(a) and \ref{fig.sig0}(b). 
This argument also explains the crossing of the dashed curves with the respective solid curves in Fig.~\ref{fig.sig0}(c).
\begin{figure}[h]
	\begin{center}
		\includegraphics[angle=-90,scale=0.31]{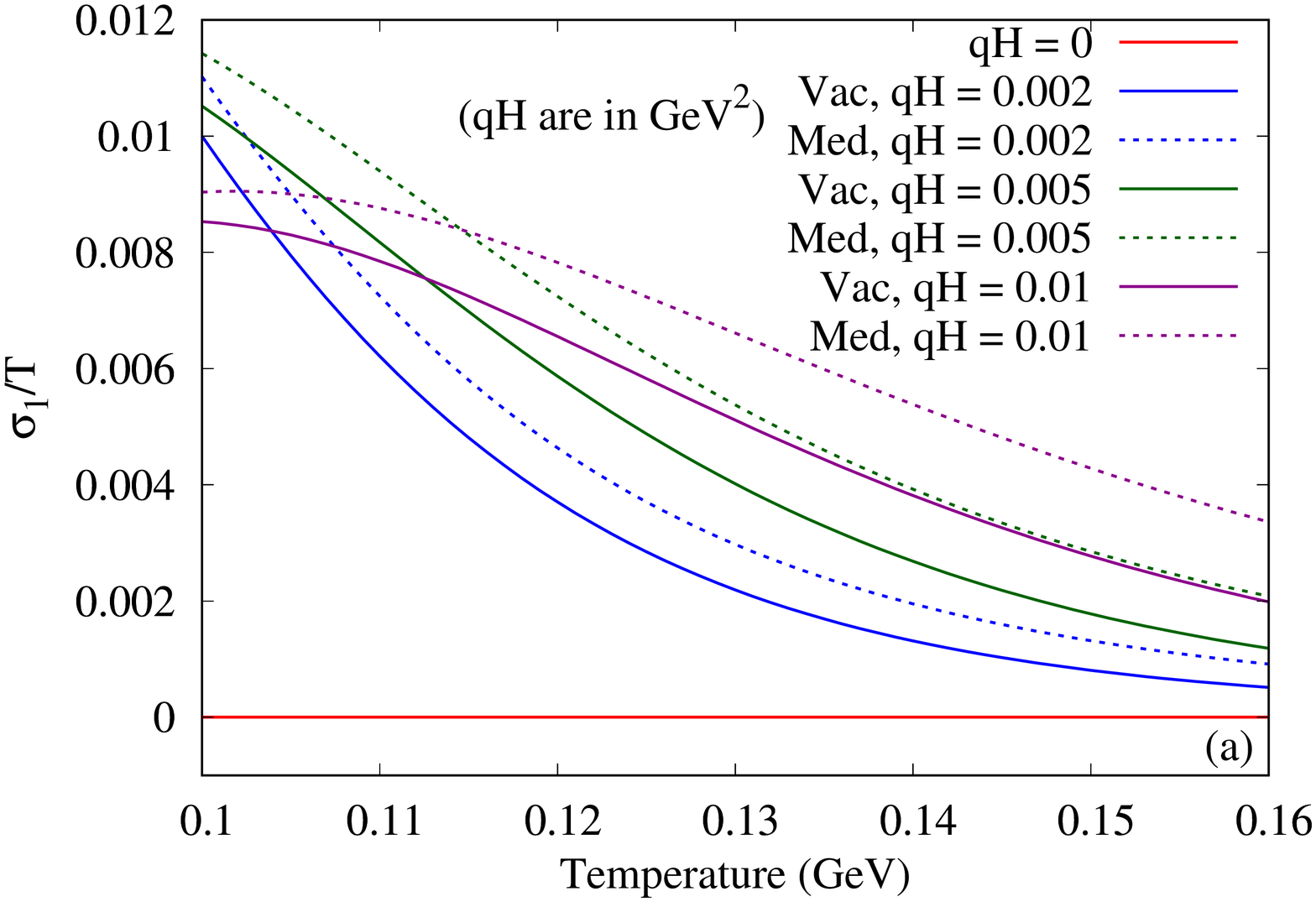}  \includegraphics[angle=-90,scale=0.31]{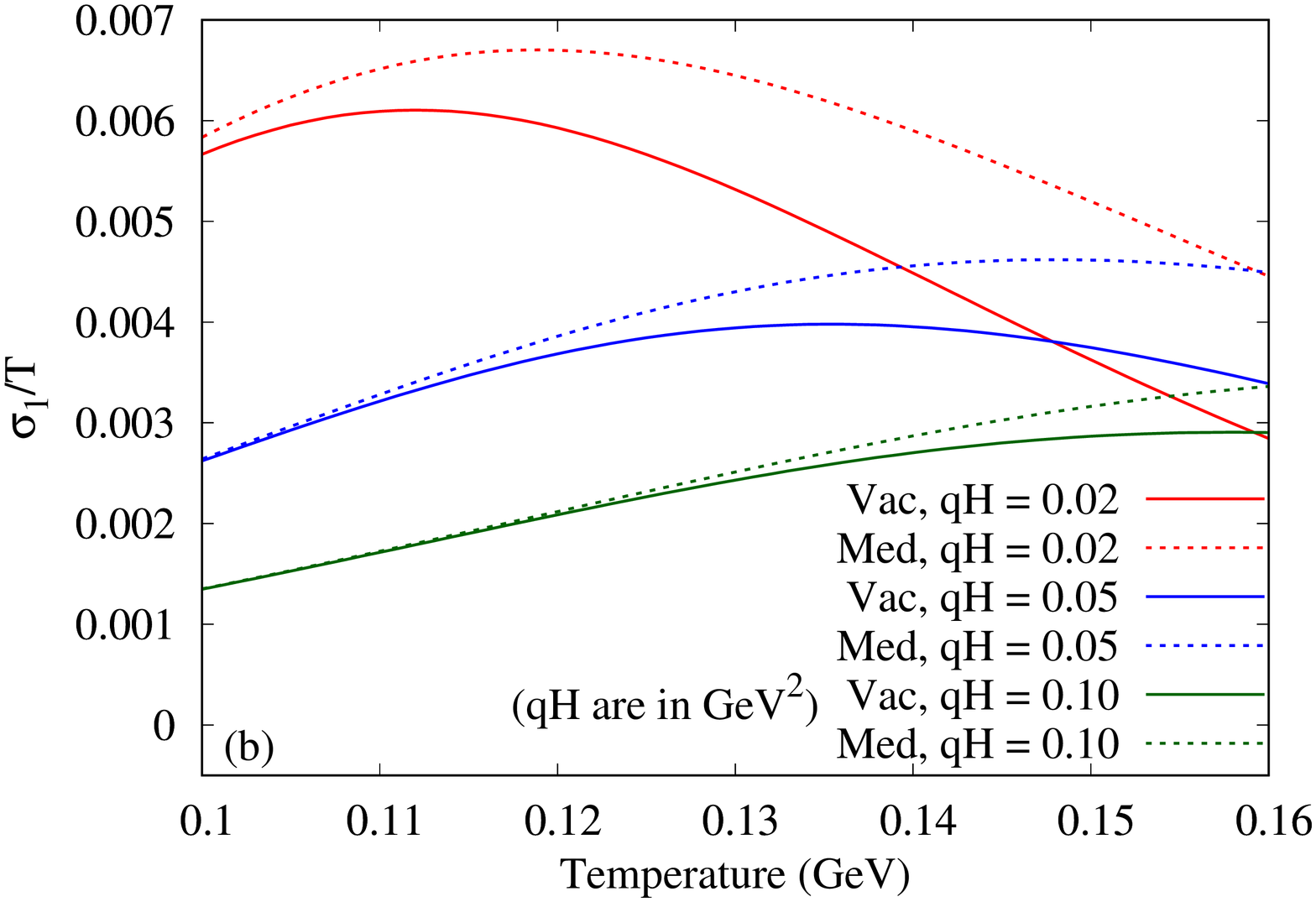}
		\includegraphics[angle=-90,scale=0.31]{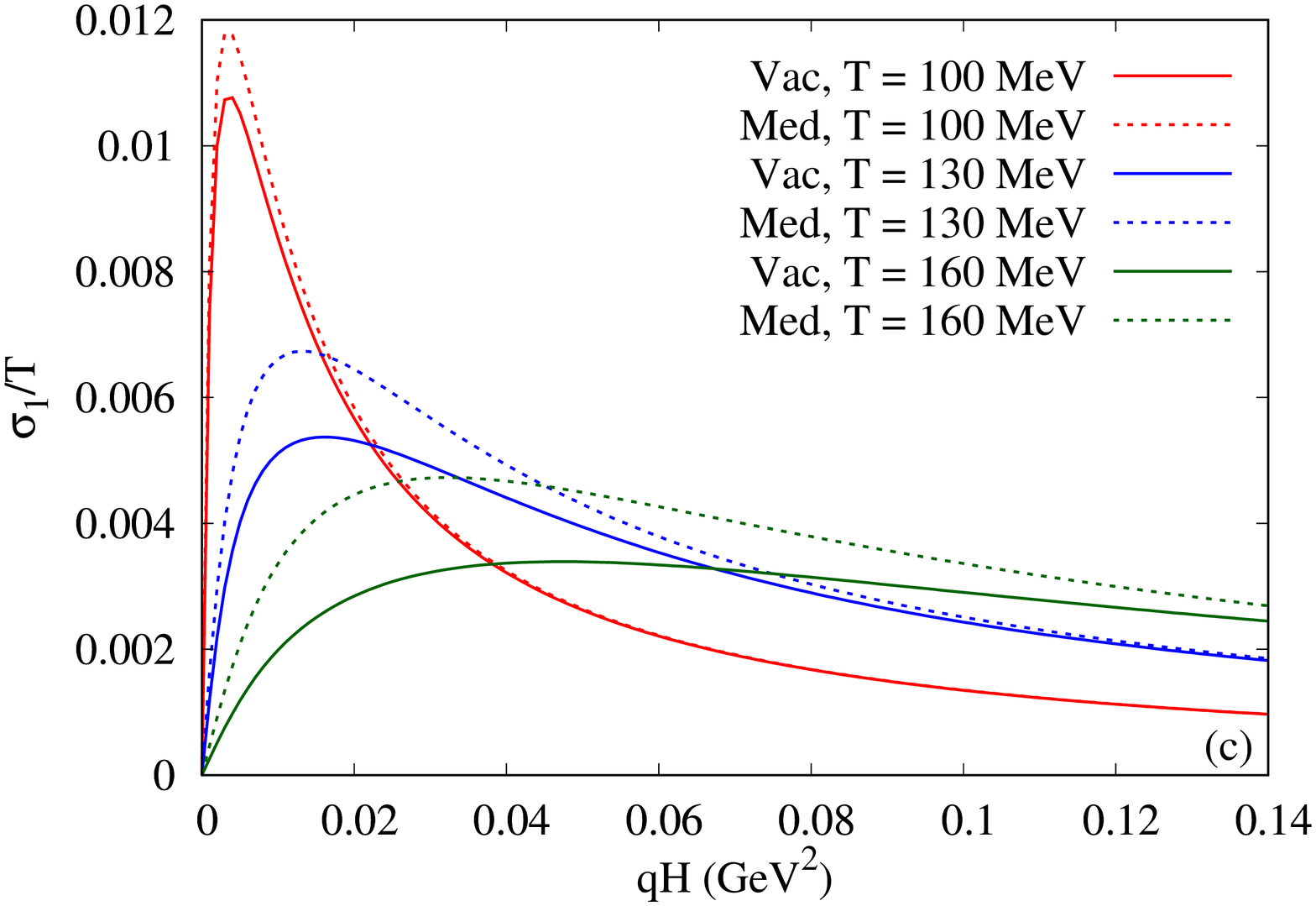}
	\end{center}
	\caption{(Color Online) The variation of $\sigma_1/T$ (a)-(b) as a function of temperature for different values of magnetic field strength and (c) as a function of magnetic field for different values of temperature. The solid and dashed curves correspond to the estimations of $\sigma_1/T$ using vacuum and in-medium cross sections respectively.}
	\label{fig.sig1}
\end{figure}

Next in Figs.~\ref{fig.sig1}(a) and \ref{fig.sig1}(b), the Hall conductivity scaled with inverse temperature ($\sigma_1/T$) has been depicted as a function of temperature 
for different values of the external magnetic field  using both the vacuum and in-medium cross sections.  
The behavior of $\sigma_1/T$ with temperature can be understood by a similar analysis 
as done in the last paragraph. We notice from Eq.~\eqref{sigma_1}, the temperature dependence of $\sigma_1/T$ approximately comes from $\sigma_1/T \sim \frac{\tau^2 T}{1+(\omega_c\tau)^2}$. 
Therefore, at lower values of the magnetic field, $\omega_c\tau \ll 1$ so that $\sigma_1/T \sim \tau^2 T \sim 1/T^5 $. 
On the other hand, at higher values of the external magnetic field ($\omega_c\tau \gg 1$), the leading temperature dependence of $\sigma_1/T$ goes as 
$\sigma_1/T \sim T $. Thus, at lower values of magnetic field, $\sigma_1/T$ decreases with temperature more rapidly than $\sigma_0/T$ whereas 
at higher values of magnetic field, we notice a linear increase of $\sigma_1/T$ with temperature. This also makes $\sigma_1/T$ to vary non-monotonically at intermediate values of the external 
magnetic field as can be noticed in Figs.~\ref{fig.sig1}(a) and \ref{fig.sig1}(b).
In Fig.~\ref{fig.sig1}(c), we have shown $\sigma_1/T$ as a function of external magnetic field for different temperatures. 
The dependence of $\sigma_1/T$ on the magnetic field goes as $\sigma_1/T \sim \frac{\omega_c}{1+(\omega_c\tau)^2}$ which is basically a 
Breit-Wigner function of the magnetic field with peak position $\sim 1/\tau\sim T^3$ and width $\sim \tau\sim 1/T^3$. The Breit-Wigner like behavior of $\sigma_1/T$ can 
be observed in Figs.~\ref{fig.sig1}(c) in which the peak position of $\sigma_1/T$ moves towards higher magnetic field values and the width increases with temperature.
As before, the effect of in-medium cross section on $\sigma_1/T$ can be understood from the $\tau$ dependence of $\sigma_1/T$,
i.e. from
$\sigma_1/T \sim \frac{\tau^2\omega_c}{1+(\omega_c\tau)^2}$ which is a monotonically increasing and saturating function of $\tau$. 
The saturation occurs in the 
low temperature (where $\tau$ is large) and low magnetic field region in which the overall $\tau$ dependence of $\sigma_1/T$ becomes weaker. 
Since the in-medium cross section yields a larger relaxation time, the use of in-medium cross section over the vacuum cross section always increases $\sigma_1/T$  
for any value of the external magnetic field as can be noticed in Figs.~\ref{fig.sig1}(a)-\ref{fig.sig1}(c). However, in the high magnetic field and low temperature 
region, due to the weakening of the $\tau$ dependence in $\sigma_1/T$, the medium effect in cross section becomes negligible as one can notice by comparing the 
separations between the dashed and solid curves of Fig.~\ref{fig.sig1}(b) (low temperature region) and Fig.~\ref{fig.sig1}(c) (high magnetic field region).
\begin{figure}[h]
	\begin{center}
		\includegraphics[angle=-90,scale=0.31]{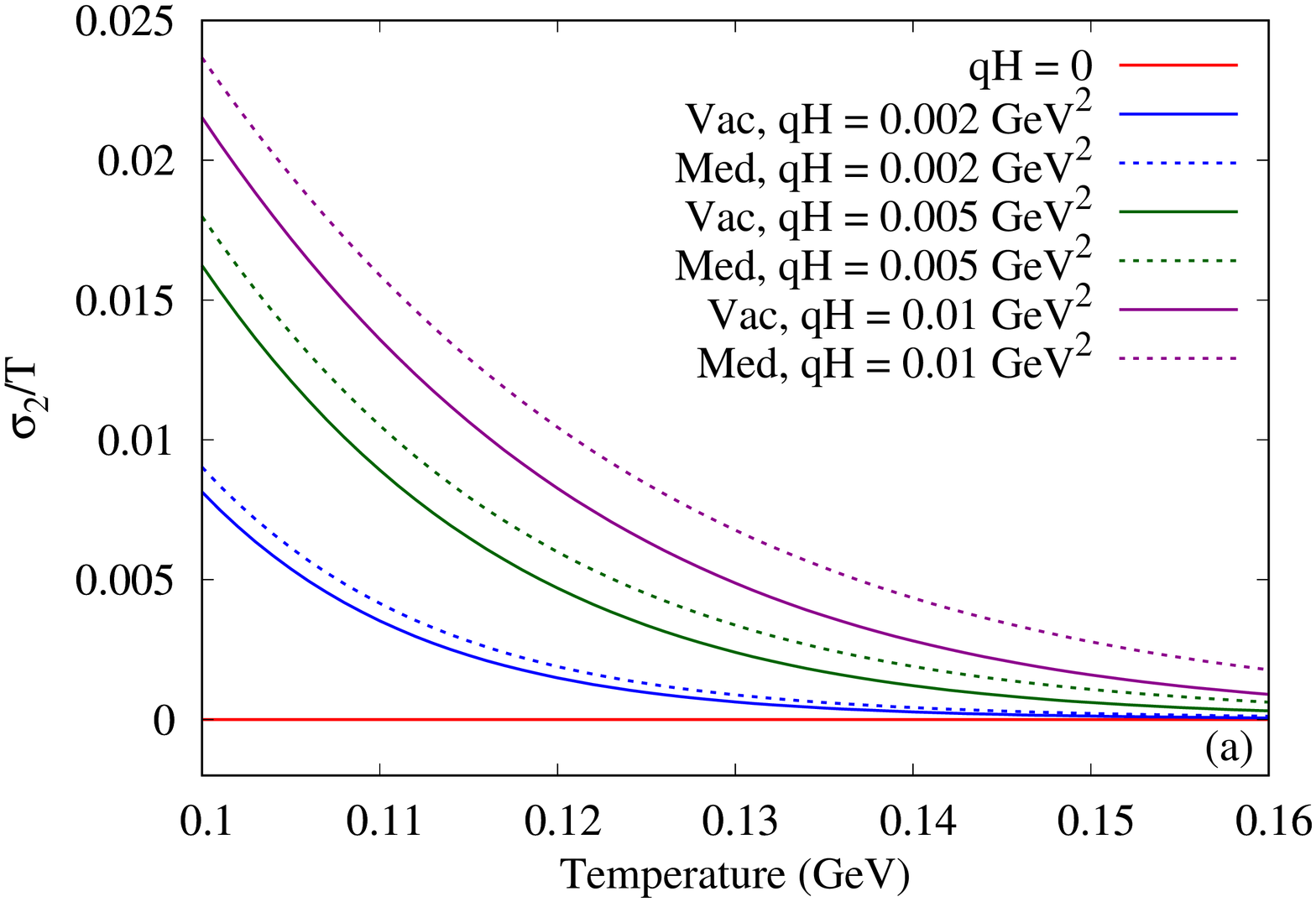}  \includegraphics[angle=-90,scale=0.31]{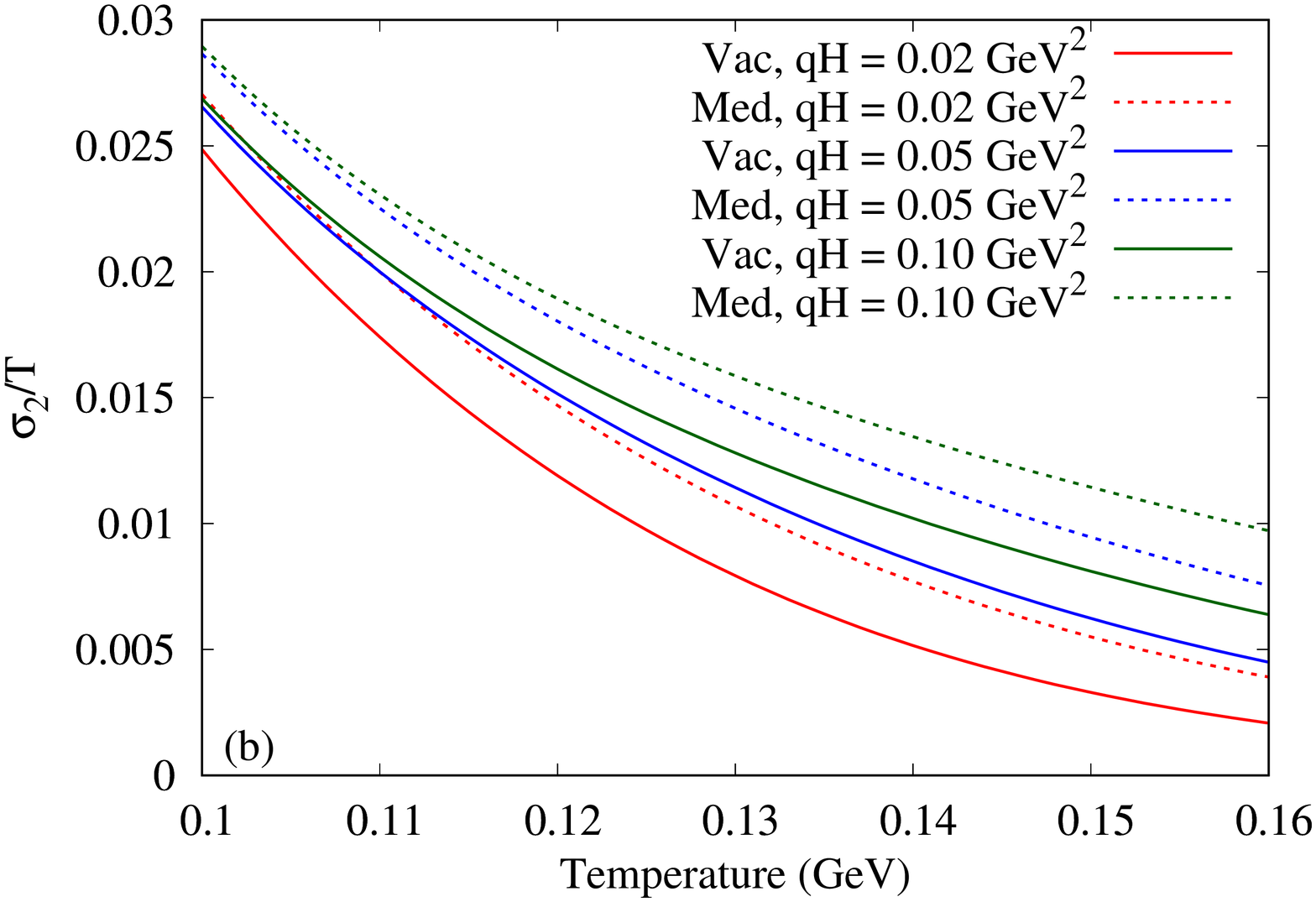}
		\includegraphics[angle=-90,scale=0.31]{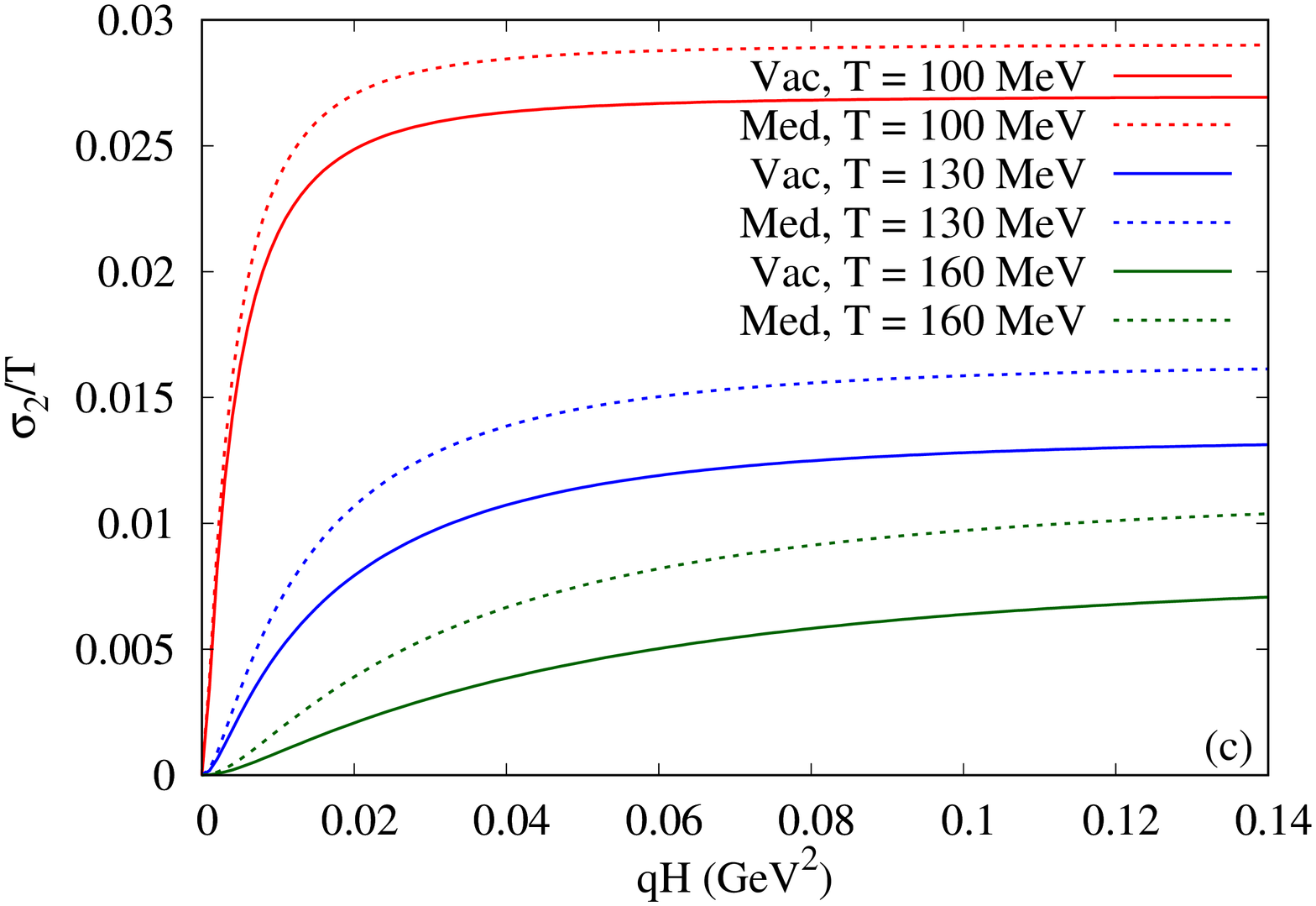}
	\end{center}
	\caption{(Color Online) The variation of $\sigma_2/T$ (a)-(b) as a function of temperature for different values of magnetic field strength and (c) as a function of magnetic field for different values of temperature. The solid and dashed curves correspond to the estimations of $\sigma_2/T$ using vacuum and in-medium cross sections respectively.}
	\label{fig.sig2}
\end{figure}

We now proceed to show the behavior of the quantity $\sigma_2/T$ as a function of temperature for different values of external 
magnetic field with both the vacuum and in-medium cross sections in Figs.~\ref{fig.sig2}(a) and \ref{fig.sig2}(b). 
The behavior of $\sigma_2/T$ with temperature can be analogously understood from Eq.~\eqref{sigma_2} in which the temperature dependence of 
$\sigma_2/T$ approximately goes as $\sigma_2/T \sim \frac{\tau^3 T}{1+(\omega_c\tau)^2}$. 
Therefore, at lower values of the magnetic field ($\omega_c\tau \ll 1$), $\sigma_2/T \sim \tau^3 T \sim 1/T^8 $. 
On the other hand, at higher values of the external magnetic field ($\omega_c\tau \gg 1$), the leading temperature dependence 
of $\sigma_2/T$ is approximately given by 
$\sigma_2/T \sim \tau \sim 1/T^3 $. Thus, $\sigma_2/T$ always decreases monotonically with the increase in temperature even more 
rapidly than $\sigma_1/T$ in all the values of the  magnetic field considered here (see Figs.~\ref{fig.sig2}(a) and 
\ref{fig.sig2}(b)).
Next, in Fig.~\ref{fig.sig2}(c), the magnetic field dependence of $\sigma_2/T$ has been depicted for different temperatures. 
$\sigma_2/T$ depends on the magnetic field as $\sigma_2/T \sim \frac{(\omega_c\tau)^2}{1+(\omega_c\tau)^2}$ which is 
a monotonically increasing and saturating function of magnetic field thus explaining the analogous behavior of the curves in the figure. 
To understand the effect of in-medium cross section on $\sigma_2/T$, we first note that the $\tau$ dependence of $\sigma_2/T$ is given by  
$\sigma_2/T \sim \frac{\tau^3 \omega_c^2}{1+(\omega_c\tau)^2}$ which is a monotonically increasing function of $\tau$ for a particular value of magnetic field. 
The rate of increase is more for higher magnetic field values. Thus, we notice, from Figs.~\ref{fig.sig2}(a)-\ref{fig.sig2}(c), that 
the use of in-medium cross section over the vacuum cross section always increases $\sigma_2/T$ for the values
of the magnetic field considered here. Moreover, at higher values 
of the external magnetic field, due to the increase of $\tau$ dependence in $\sigma_2/T$, the medium effects become more significant as can be observed by comparing the 
separations between the dashed and solid curves of Fig.~\ref{fig.sig2}(a)-\ref{fig.sig2}(c).
\begin{figure}[h]
	\begin{center}
		\includegraphics[angle=-90,scale=0.31]{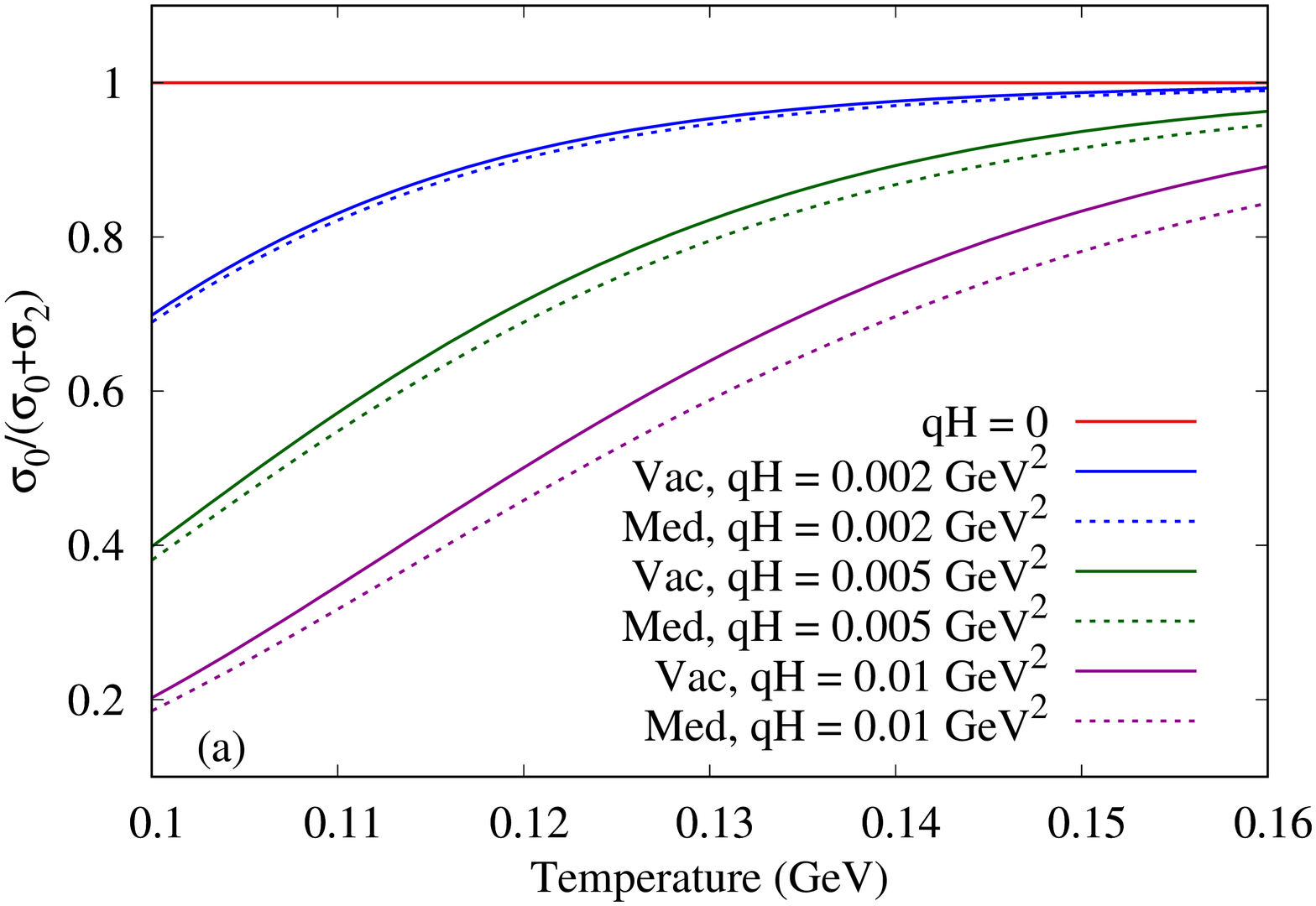}  \includegraphics[angle=-90,scale=0.31]{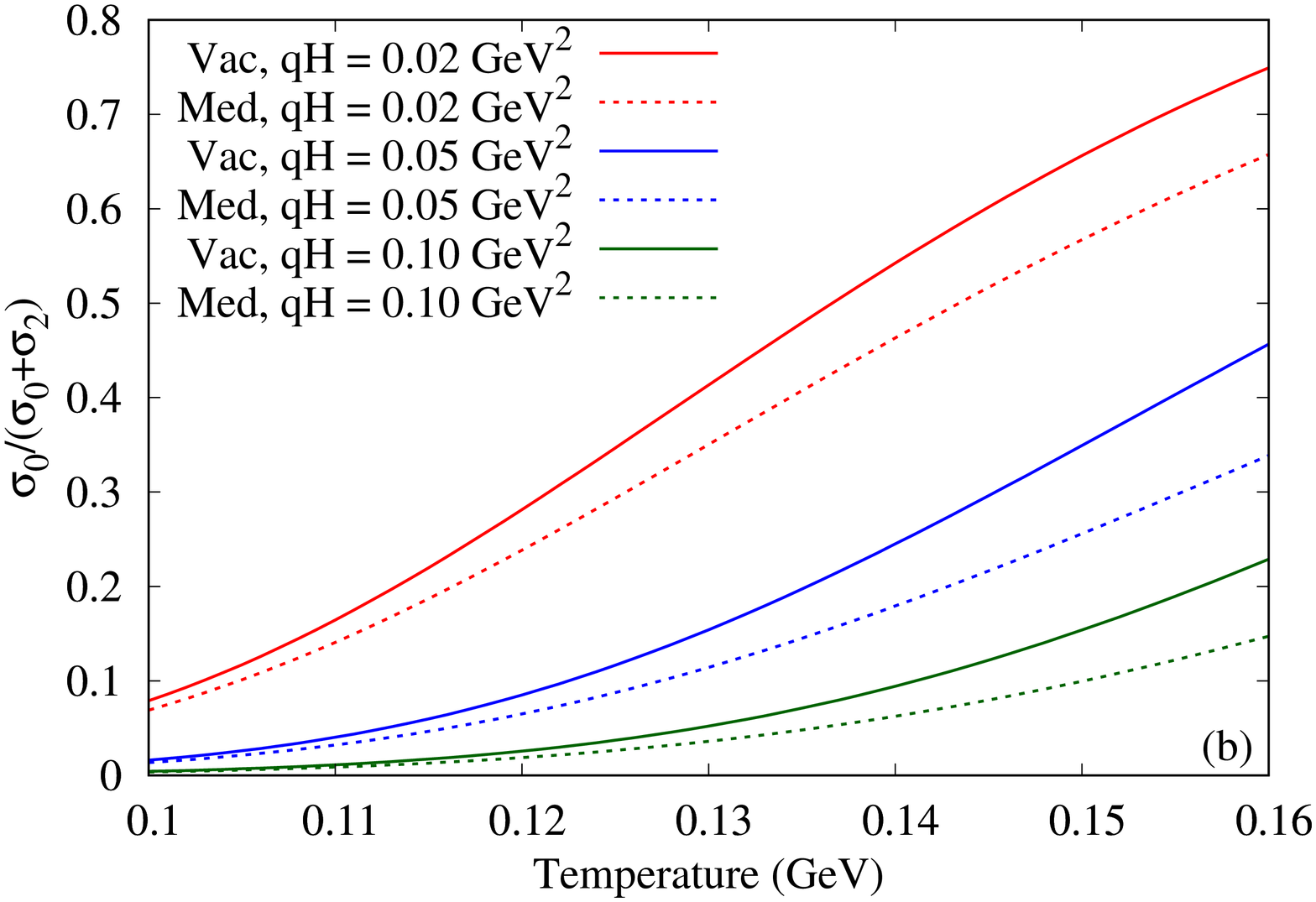}
		\includegraphics[angle=-90,scale=0.31]{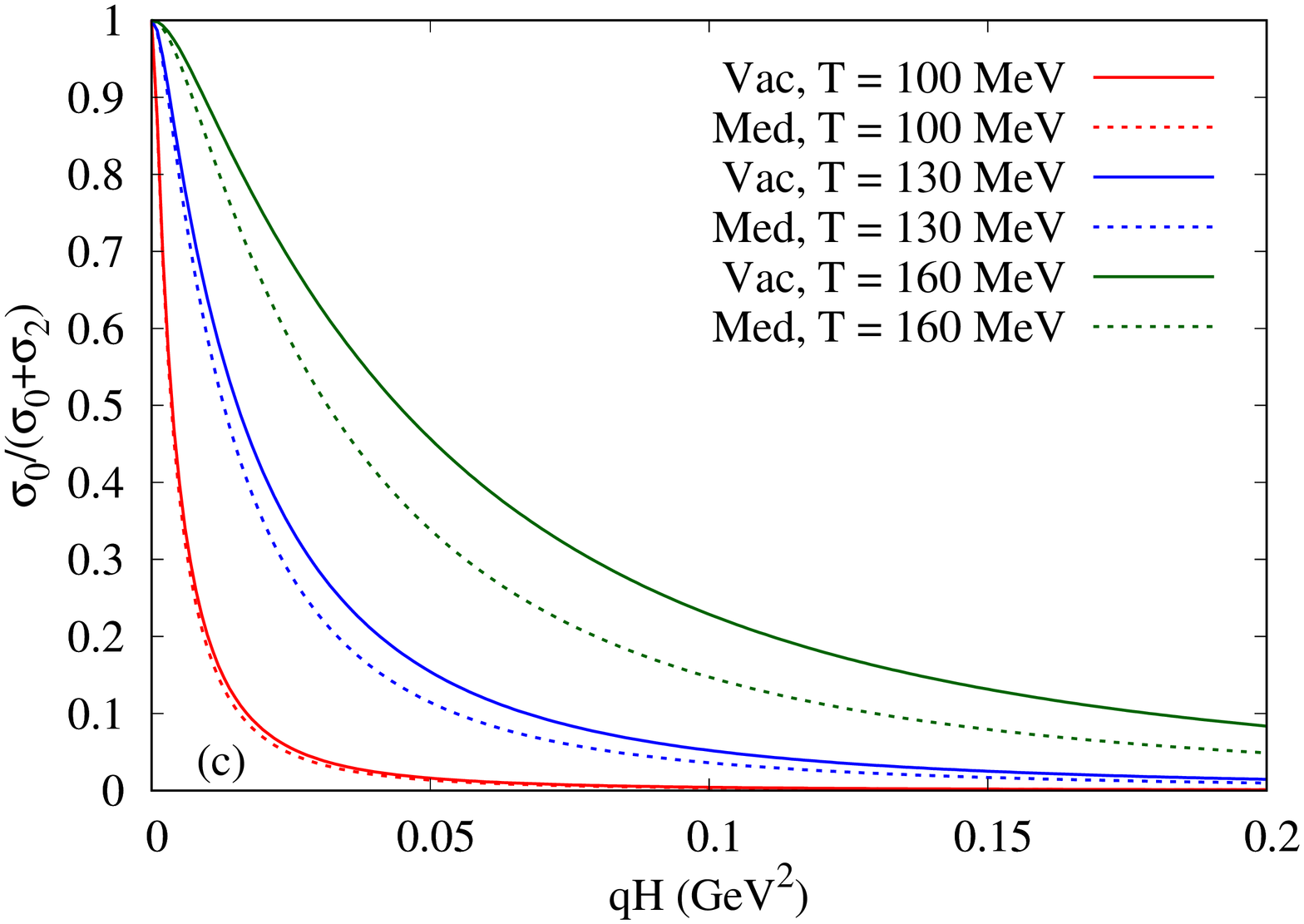}
	\end{center}
	\caption{(Color Online) The variation of the anisotropy measure $\frac{\sigma_0}{\sigma_0+\sigma_2}$ (a)-(b) as a function of temperature for different 
		values of magnetic field strength and (c) as a function of magnetic field for different values of temperature. The solid and dashed curves correspond 
		to the estimations of $\frac{\sigma_0}{\sigma_0+\sigma_2}$ using vacuum and in-medium cross sections respectively.}
	\label{fig.ratio}
\end{figure}
\begin{figure}[h]
	\begin{center}
		\includegraphics[angle=-90,scale=0.31]{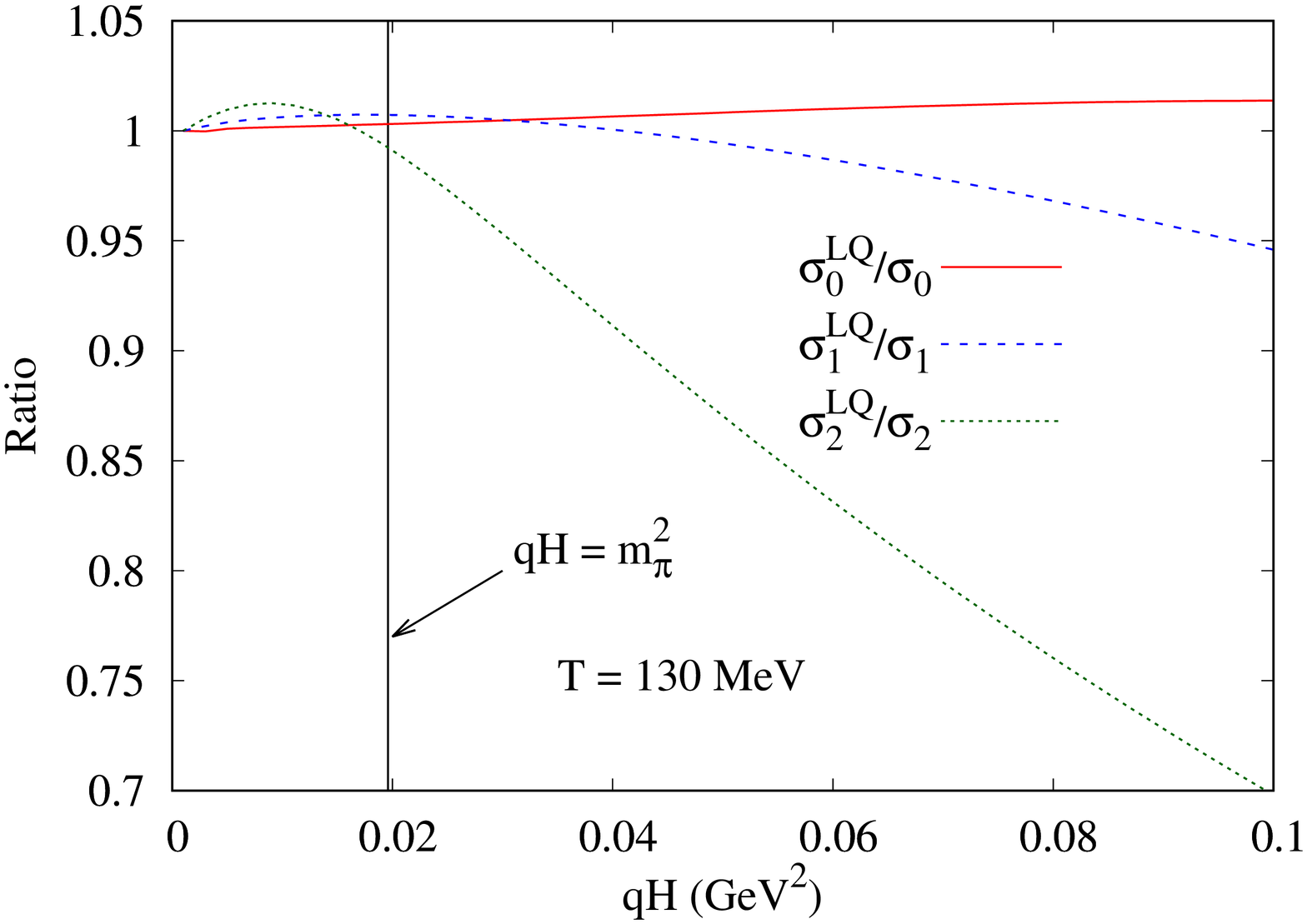} 
	\end{center}
	\caption{(Color Online) The ratio $\sigma_n^\text{LQ} / \sigma_n$ as a function of external magnetic field at $T=130$ MeV. The solid-black vertical line corresponds to $qH = m^2$. Upto 300 Landau levels are taken into consideration.}
	\label{fig.ratio.2}
\end{figure}

Finally, we note that, the normalized ratio $\frac{\sigma_0}{\sigma_0+\sigma_2}$ could be a measure of anisotropy brought in by the external magnetic field since 
the quantity $\sigma_0+\sigma_2$ is the electrical conductivity in absence of magnetic field. We therefore plot $\frac{\sigma_0}{\sigma_0+\sigma_2}$ as 
a function of temperature and magnetic field in Figs.~\ref{fig.ratio}(a)-\ref{fig.ratio}(c). With the increase in temperature, the ratio increases towards its 
asymptotic value $1$ whereas with the increase in magnetic field, the ratio rapidly decreases from $1$. Physically it corresponds to the fact that, the magnetic 
field tries to bring anisotropy in the medium whereas the thermal fluctuation tries to diminish it. Moreover, comparing the solid and dashed curves in 
Figs.~\ref{fig.ratio}(a)-\ref{fig.ratio}(c), we find that the use of medium effects in the cross section makes the system more anisotropic in presence 
of external magnetic field.

We have already mentioned in Sec.~\ref{sec.sigma} that we are neglecting the LQ of the charged pion dispersion relation (see Eq.~\eqref{eq.LQ}) while calculating the conductivities. However, to check the validity of this continuum approximation, let us now calculate the conductivities incorporating the LQ of pion transverse momentum. 
To a first approximation, the LQ can be incorporated in the final expression of the conductivities in Eq.~\eqref{sigma_n} by the following replacements:
\begin{eqnarray}
\omega_p &\to& \omega_{pl} = \sqrt{p_z^2 + (2l+1)qH + m^2}, \\ 
|\vec{p}| &\to& \sqrt{p_z^2 + (2l+1)qH}, \\
\int \!\!\! \frac{d^3 p}{(2\pi)^3}  &\to& \frac{qH}{2\pi} \sum_{l=0}^{\infty} \int_{-\infty}^{\infty}\frac{dp_z}{2\pi},
\end{eqnarray}
so that the conductivities with LQ becomes
\begin{eqnarray}
\sigma_n^\text{LQ} = \frac{g q^2}{3T} \frac{qH}{2\pi} \sum_{l=0}^{\infty} \int_{-\infty}^{\infty}\frac{dp_z}{2\pi} \frac{p_z^2 + (2l+1)qH}{\omega_{pl}^2} \frac{\tau (\omega_{c,l} \tau)^n}{1+ (\omega_{c,l} \tau)^2} f_0(\omega_{pl})\SB{1+f_0(\omega_{pl})} \label{sigma_n.lq}
~~~;~~~ n=0,1,2,
\end{eqnarray}
where $\omega_{c,l} = qH/\omega_{pl}$. In Fig.~\ref{fig.ratio.2}, we have shown the ratio $\sigma_n^\text{LQ} / \sigma_n$ as a function of external magnetic field at $T=130$ MeV. From the figure, we can see that in the low magnetic field region ($qH \precsim m^2$), the ratios are almost unity which imply that the use of continuum approximation is well justified in the weak field region. However, for higher magnetic field values, the continuum approximation breaks down and the LQ becomes important. For example, at $qH=0.10$ GeV$^2$, LQ modifies the values of $\sigma_0$ and $\sigma_1$ by less than 5\% whereas the change in $\sigma_2$ is about 30\%. Therefore, even if we have shown numerical results for a wider range of magnetic field values ($0 \le qH \le 0.1$ GeV$^2$) neglecting the LQ, our results are strictly valid for the weak magnetic field ($0 \le qH < m^2$) likely  to be realised in the hadronic phase of HIC.

\begin{figure}[h]
	\begin{center}
		\includegraphics[angle=-90,scale=0.31]{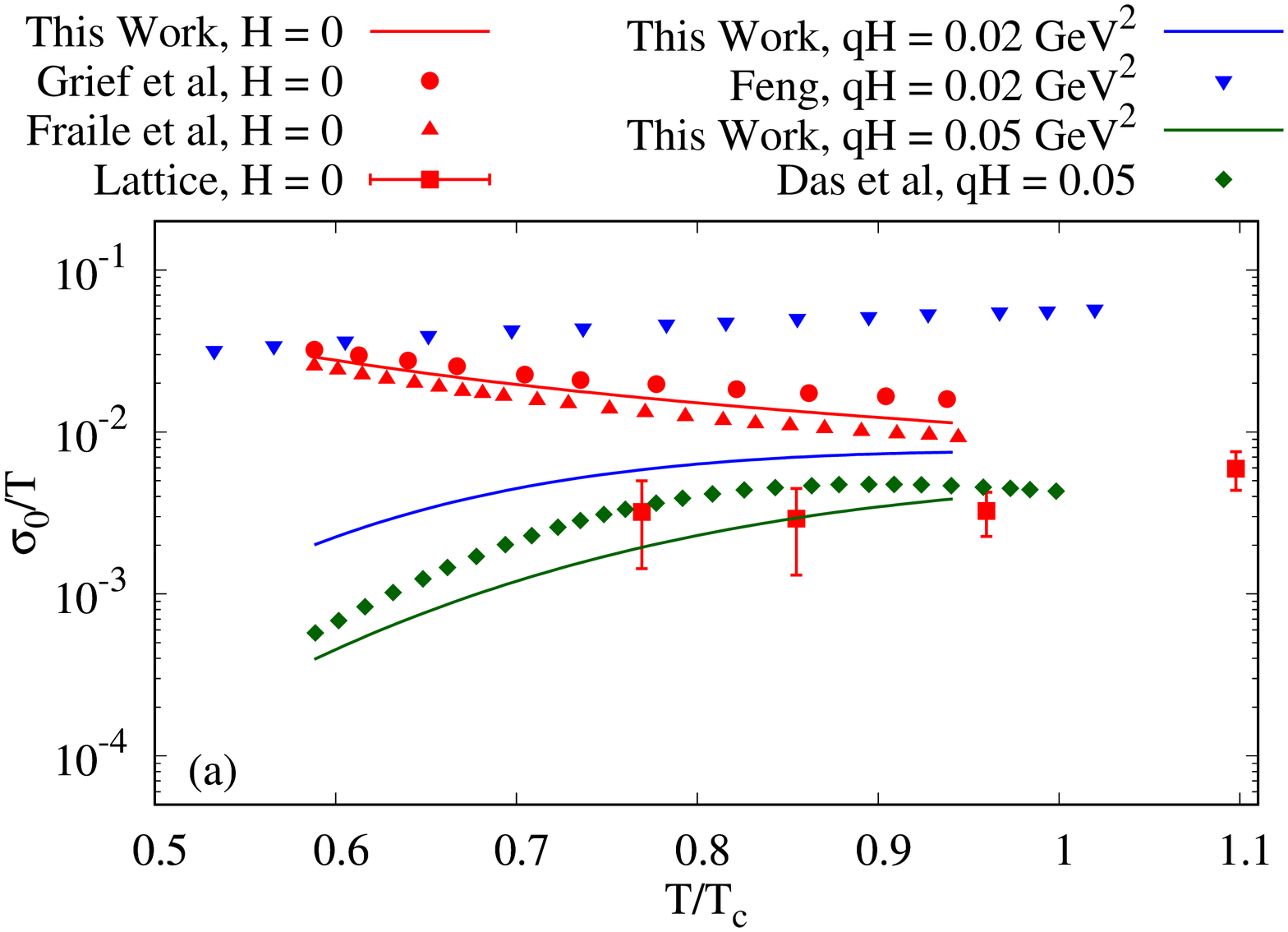}  \includegraphics[angle=-90,scale=0.31]{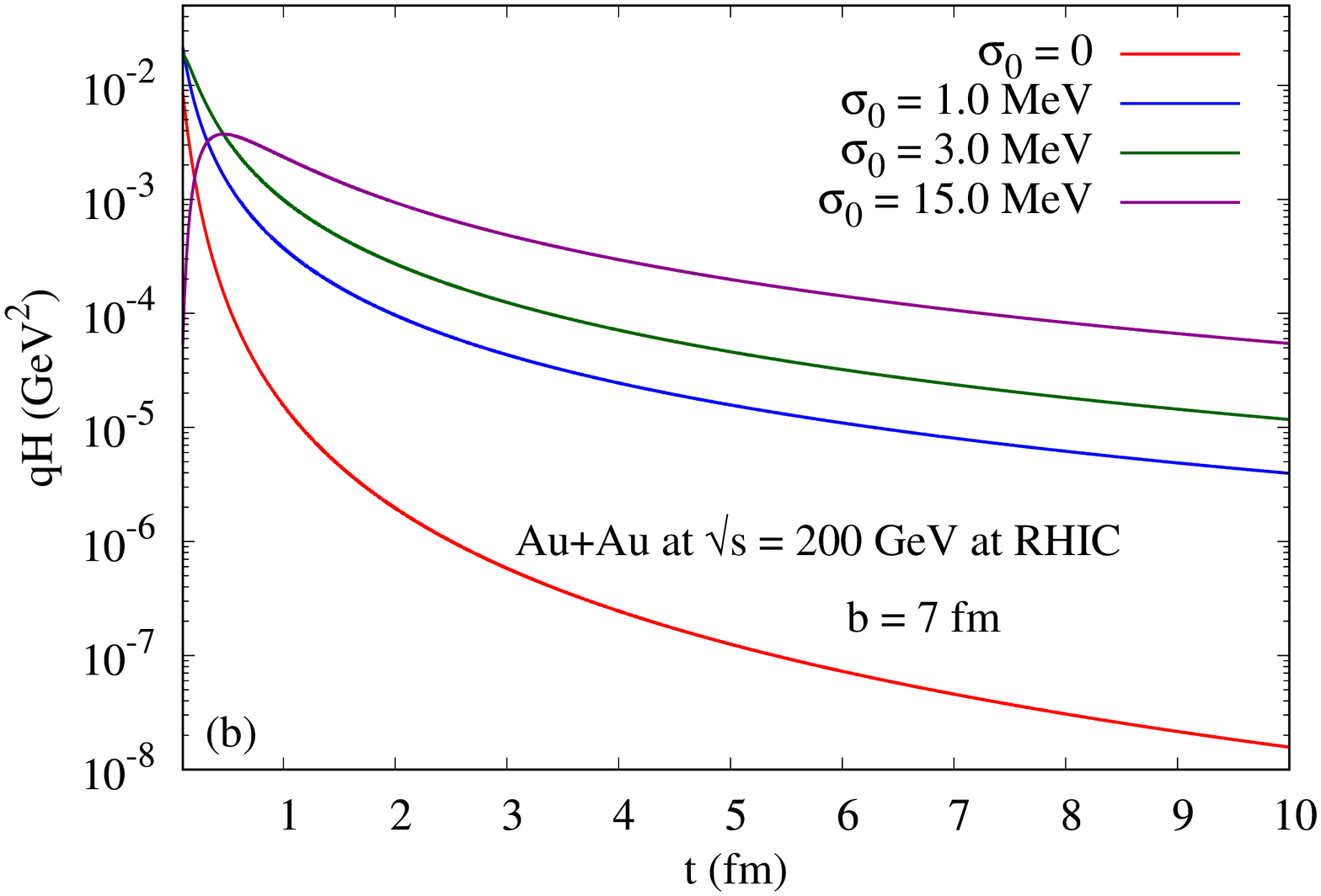} 
	\end{center}
	\caption{(Color Online) (a) The comparison of $\sigma_0/T$ at zero magnetic field with Grief et al~\cite{Greif:2016skc}, Fraile et al~\cite{FernandezFraile:2005ka} and Lattice QCD calculation~\cite{Amato:2013naa} and at non-zero magnetic field with Feng~\cite{Feng:2017tsh} and Das et al~\cite{Das:2019wjg}. (b) The decay of maximum magnetic field value in peripheral Au+Au collision at RHIC for different values of electrical conductivities. }
	\label{fig.compare}
\end{figure}

In Fig.~\ref{fig.compare}(a), we have made a comparison of the electrical conductivity obtained in this work with the other available estimations in the literature. We see that, our estimation of electrical conductivity at zero-magnetic field agrees well with the earlier estimations by Grief et al~\cite{Greif:2016skc} and Fraile et al~\cite{FernandezFraile:2005ka} whereas it does not agree well with the Lattice QCD estimation~\cite{Amato:2013naa}. Also, our result at $qH=0.02$ GeV$^2$ is lower than the values obtained by Feng.~\cite{Feng:2017tsh} for a system of relativistic quark-gluon gas. This is expected as the conductivity of QGP is much larger than of hadron gas. Finally, our result at $qH=0.05$ GeV$^2$ is in good qualitative and quantitative agreement with the result of Das et al~\cite{Das:2019wjg} calculated for hadron resonance gas.

We have already mentioned in Sec.~\ref{sec.intro},  that a sufficiently high value of electrical conductivity of the medium can sustain the rapidly decaying magnetic field in a HIC~\cite{Gursoy:2014aka,Skokov:2009qp,Tuchin:2010vs,Tuchin:2013ie}. To see how our estimated electrical conductivity (for a system of pion gas) modifies the decay of the magnetic field in HIC, we have calculated the time ($t$) dependence of the maximum value of magnetic field for peripheral Au+Au collision at RHIC energy ($\sqrt{s}$=200 GeV) using the simplified expression used by Tuchin~\cite{Tuchin:2013ie} for a static medium. In Fig.~\ref{fig.compare}(b), we have plotted the decay of the maximum magnetic field value in peripheral Au+Au collision at RHIC for different values of electrical conductivities. In our calculation, we have obtained maximum value of $\sigma_0$ as 3 MeV whereas in a QGP medium it has typical value of $\simeq15$ MeV~\cite{Ding:2010ga,Aarts:2007wj,Gursoy:2014aka,Tuchin:2010vs}. From the figure, it can be noticed that, for a constant $\sigma_0=15$ MeV throughout the evolution, a magnetic field value of the order of $10^{-4}$ GeV$^2$ is sustained even at $t=10$ fm. But if we consider constant $\sigma_0=1-3$ MeV (as obtained in the current work for a pion gas) throughout the evolution, the sustained value of magnetic field at  $t=10$ fm is of the order of $10^{-5}$ GeV$^2$. In reality, electrical conductivity is not expected to be constant throughout the evolution.  In the early stage (QGP phase), $\sigma_0$ will be large ($\sim 15$ MeV) and in the later stages (hadronic phase) $\sigma_0$ will be small ($\sim 5$ MeV). Therefore the time evolution of the actual magnetic field value is expected to lie in between the violet and green curves in the figure. Moreover, in Fig.~\ref{fig.compare}(b), we have considered a medium with no hydrodynamic expansion for the estimation of the decay of magnetic field. For an expanding medium (which is the more realistic scenario for HIC), the magnetic field will sustain for a longer period as shown in Ref.~\cite{Tuchin:2013ie}. Thus, we can conclude that a weak magnetic field can be present in the later stages of HIC and could be phenomenologically relevant.


\section{Summary \& Conclusions} \label{sec.summay}
We have evaluated the conductivity tensor using the Boltzmann transport equation in a magnetic field and hence evaluated the the electrical conductivity, Hall conductivity and $\sigma_2$ for a system consisting of a pion gas. The information pertaining to the pion gas enters through the relaxation time into the expression of the three conductivities. The $\pi\pi$ cross section has been calculated in a thermal medium using the real time formalism of finite temperature field theory. We have shown the variation of these three conductivities with temperature for different values of the  magnetic field. It is observed that electrical conductivity and Hall conductivity are very sensitive to the magnetic field strength and the in-medium cross sections. Moreover, as we have not considered the LQ in the dispersion relation of charged pions, our results are more accurate in the low magnetic field values ($qH \lesssim m^2$), which is the realistic scenario for the later stages of HICs.

Both the electrical and Hall conductivities have been found to increase with temperature for a given value of the magnetic field when the in-medium cross-section is used. For a given temperature there is no appreciable change (except at lower $B$) in the electrical conductivity with the magnetic field when medium dependent cross section is used. A more detailed observation shows a monotonically increasing trend of electrical conductivity with the increase in temperature at higher values of the magnetic field. However, for a given temperature the conductivity is found to decrease monotonically as a function of the magnetic field. In the case of Hall conductivity it is found that at lower values of the magnetic field, it decreases with the increase in temperature more rapidly than the electrical conductivity, whereas at higher values of the magnetic field, a linear increase of the Hall conductivity with the	temperature is observed. For a given temperature as long as it is low we see a Breit-Wigner type structure in the Hall conductivity as a function of the magnetic field. This structure disappears and tends to saturate at higher temperature. This behaviour can be attributed to the substantial spectral broadening of the exchanged particle at high temperature.

The electrical conductivity obtained in this work is shown to have both qualitative and quantitative agreement with earlier estimates available in the literature. Moreover, the calculated electrical conductivity is shown to be sufficient for causing a significant delay in the decay of external magnetic field in HIC. This leads to the conclusion that, weak magnetic field can be present in the later stage of HIC (in hadronic phase) and could be phenomenologically relevant.

Finally, we should mention that we have included only pions in our calculations as they are more abundantly produced in the temperature range achievable in HICs. Hadrons heavier than the pion can, in principle, be included as is done in Ref.~\cite{Das:2019wjg} using a constant cross section. However, in the present formalism it will be extremely non-trivial to solve the coupled transport equations as well as to calculate a plethora of cross sections due to the inclusion of multiple species.

\section*{Acknowledgements}
The authors are funded by the Department of Atomic Energy (DAE), Government of India.

\bibliographystyle{apsrev4-1}
\bibliography{pallavi}

\end{document}